\documentclass[a4paper,twocolumn,11pt,submission]{quantumarticle}
\pdfoutput=1

\usepackage{braket}
\usepackage{graphicx}% Include figure files
\usepackage{dcolumn}% Align table columns on decimal point
\usepackage{bm}% bold math
\usepackage{hyperref}% add hypertext capabilities
\usepackage{float}
\usepackage{amsthm}
\newtheorem{lemma}{Lemma}

\usepackage[backend=bibtex]{biblatex}
\usepackage{xcolor}
\usepackage{comment}
\usepackage{mathtools}

\usepackage[ruled,lined]{algorithm2e}
\usepackage{algpseudocode}

\usepackage{dsfont}

\usepackage{svg}
\usepackage{makecell}

\usepackage[normalem]{ulem}

\usepackage{tikz}
\usetikzlibrary{shapes, arrows.meta, positioning}

\definecolor{gold}{rgb}{0.83, 0.69, 0.22}

\DeclareMathOperator*{\argmin}{argmin}

\begin{document}

\title{A Joint Quantum Computing, Neural Network and Embedding Theory Approach for the Derivation of the Universal Functional}

\author{Martin J. Uttendorfer} \email{martin.uttendorfer@dlr.de}
\affiliation{Institute for Frontier Materials on Earth and in Space, German Aerospace Center (DLR), 51170 Cologne, Germany}
\author{Daniel Barragan-Yani}
\affiliation{Institute for Frontier Materials on Earth and in Space, German Aerospace Center (DLR), 51170 Cologne, Germany}
\author{Matthias Sperl}
\affiliation{Institute for Frontier Materials on Earth and in Space, German Aerospace Center (DLR), 51170 Cologne, Germany}
\author{Marc Landmann}
\affiliation{Institute for Frontier Materials on Earth and in Space, German Aerospace Center (DLR), 51170 Cologne, Germany}

\date{August 7, 2026}

\begin{abstract}
We introduce a novel approach that exploits the intersection of quantum computing, machine learning and reduced density matrix functional theory to leverage the potential of quantum computing to improve simulations of interacting quantum particles. Our method focuses on obtaining the universal functional using a deep neural network trained with quantum algorithms. In addition, we use density matrix embedding theory to strengthen our approach by substantially expanding the space of Hamiltonians for which the obtained functional can be applied without the need for additional quantum resources. Since the obtained universal functional can be reused for any system where the interactions within the embedded fragment are identical, our work demonstrates a way to potentially achieve a \textit{cumulative} quantum advantage within quantum computing applications for quantum chemistry and condensed matter physics.
\end{abstract}

\maketitle

\section{\label{sec:Introduction}Introduction}

Density functional theory (DFT)\cite{Hohenberg1964} has become a universal tool in quantum chemistry and condensed matter physics. Although proven highly successful for a wide range of electronic-structure problems, DFT lacks an accurate intrinsic description of strong correlation \cite{Teale2022}. In this regard, the conceptual quantum-computer scaling advantage in treating strongly interacting domains by fully-correlated quantum-chemical methods~\cite{Elfving2020, Chee2024, Liu2021} suggests the idea of combining quantum computing (QPUs) and functional theory at a fundamental level. Initial steps in this direction include quantum-enhanced approaches to solving the Kohn-Sham equations~\cite{Senjean2023} and quantum-assisted functional optimization~\cite{Sheridan2024,Koridon24}. Building upon variational quantum algorithms (VQAs), several groups have implemented Levy-Lieb's constrained search \cite{Levy1979} on quantum hardware. For example, Pemmaraju et al.~\cite{Pemmaraju2022} embedded the constrained-search formalism into DFT applications, while Schade et al.~\cite{Schade2022} extended this approach to reduced density matrix functional theory (RDM-FT), enabling the treatment of larger systems through the adaptive cluster approximation \cite{Schade2018}.

% Density functional theory (DFT), the foundations of which were laid out in the seminal 1964 work by Hohenberg and Kohn \cite{Hohenberg1964}, has become a universal tool in quantum chemistry as well as condensed matter physics. While DFT has proven highly successful for a wide range of electronic-structure problems, DFT lacks a universal intrinsic description of strong correlation \cite{Teale2022}. In contrast, the conceptual quantum-computer immanent scaling advantage in treating strongly interacting domains by full-correlated quantum-chemical methods~\cite{Elfving2020, Chee2024, Liu2021} suggests the idea of combining quantum processing units (QPUs) and functional theory at a fundamental level. Initial steps in this direction include quantum-enhanced approaches to solving the Kohn-Sham equations~\cite{Senjean2023} and quantum-assisted functional optimization~\cite{Sheridan2024,Koridon24}. Building upon variational quantum algorithms (VQAs), several groups have implemented Levy-Lieb's constrained search \cite{Levy1979} on quantum hardware. For example, Pemmaraju et al.~\cite{Pemmaraju2022} embedded the constrained-search formalism into DFT applications, while Schade et al.~\cite{Schade2022} extended this approach to reduced density matrix functional theory (RDM-FT), enabling the treatment of larger systems through the adaptive cluster approximation \cite{Schade2018}. 

In the present work, machine learning (ML) methods are used to complement quantum algorithms used in the context of RDM-FT. Our ML-enhanced quantum framework is, in principle, agnostic to the specific quantum algorithm employed, e.g., variational quantum eigensolver (VQE) and its variants, quantum phase estimation (QPE), quantum subspace expansion (QSE) and quantum-selected configuration interaction (QSCI) \cite{Kanno2023, Nakagawa2024}, as well as independent of the quantum hardware utilized. Furthermore, our framework can potentially offer significant advantages when scaling to larger embedded systems  as density matrix embedding theories main scaling component is a relatively cheap mean-field method. In addition, based on Gilbert's work \cite{Gilbert1975} and subsequent refinements \cite{Piris2014, Pernal2015, Giesbertz2019, Liebert20232}, RDM-FT offers more general  functionals than the regular DFT formalism. Specifically, by fixing only the interaction term of the many-body Hamiltonian, the same RDM-functional is useful to solve more systems than a functional with the kinetic hopping term being fixed. Closely related to our workflow, and with a focus on efficient data generation, Baker and Poulin \cite{Baker2020} propose a scheme to learn the density functional for Kohn-Sham-DFT using quantum hardware, specifically using QPE.

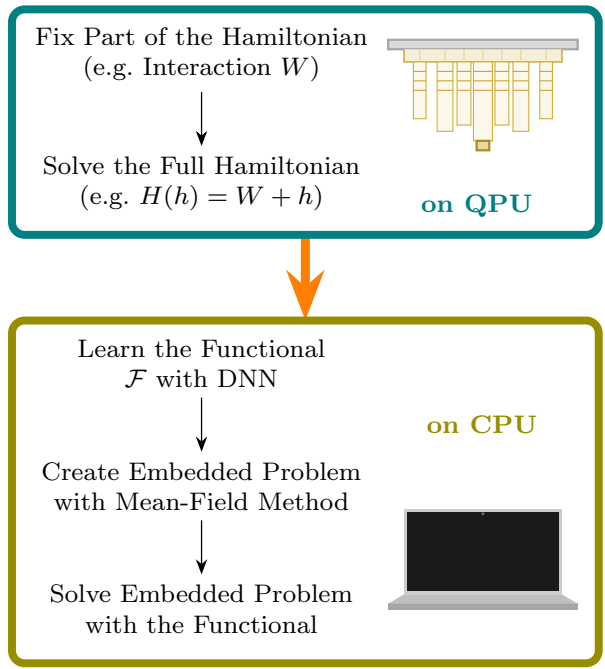
\begin{figure}[H]
    \centering
    \scalebox{1.2}{
   \begin{tikzpicture}[
    node distance=0.2cm,
    every node/.style={rounded corners, align=center, font=\footnotesize},
    arrow/.style={-{Stealth}}
]

    \node (FirstNode) {\tiny Fix Part of the Hamiltonian \\ \tiny (e.g.~Interaction $W$)};
    %\draw[arrow] (FirstNode) -- (ThirdNode);

    \node (ThirdNode) [below=of FirstNode, yshift=-0.4cm] {\tiny Solve the Full Hamiltonian \\ \tiny (e.g.~$H(h)=W+h$)};
    \draw[arrow] (FirstNode) -- (ThirdNode);

    \node (FourthNode) [below=of ThirdNode, yshift=-0.4cm] {\tiny Learn the Functional \\ \tiny $\mathcal{F}$ with DNN};
    \draw[arrow] (ThirdNode) -- (FourthNode);
    
    \draw[arrow, line width=1mm, color=orange] (1.15,-3.3) -- (1.15,-4.25);
    
    \node (FifthNode) [below=of FourthNode, yshift=-1cm] {\tiny Create Embedded Problem \\ \tiny with Mean-Field Method};
    %\draw[arrow] (FourthNode) -- (FifthNode);
    
    \node (SixthNode) [below=of FifthNode, yshift=-0.4cm] {\tiny Solve Embedded Problem \\ \tiny with the Functional};
    \draw[arrow] (FifthNode) -- (SixthNode);
    
    \draw[draw=teal, line width=2.5pt, rounded corners] (-2.1, .5) rectangle (4.4,-3.3);
    \node at (3.1,-2.2) {\color{teal} \textbf{\tiny Execute Once} \\ \color{teal} \textbf{\tiny Including} \\ \color{teal} \textbf{\tiny Use of QPUs} };
    
    \draw[draw=olive, line width=2.5pt, rounded corners] (-2.1, -4.25) rectangle (4.4,-6.75);
    \node at (3.1,-4.95) {\color{olive} \textbf{\tiny Reuse Many} \\ \color{olive} \textbf{\tiny Times Only} \\ \color{olive} \textbf{\tiny on CPUs} };

    \draw [draw, rounded corners, align=center, fill=teal, opacity=0.25, font=\footnotesize, line width=0pt](-1.9, -1.05) rectangle (1.9,-1.8);

    \node at (3.1,-.7) {\scalebox{.3}{
\begin{tikzpicture}
% Top mounting structure
\filldraw[fill=gray!30, draw=gray!70, line width=2pt, sharp corners] 
    (-3, 4.5) rectangle (3, 4.2);

% Main support structure
\filldraw[fill=gold!20, draw=gold!100!black, line width=1.5pt, sharp corners] 
    (-2.5, 4.2) -- (-2.5, 3.8) -- (2.5, 3.8) -- (2.5, 4.2);

% The iconic golden hanging cylinders in a chandelier pattern
\foreach \x/\y/\h/\r in {
    0/3.8/2.5/0.3,           % Center - longest
    -1.2/3.8/2.2/0.25,       % Inner left
    1.2/3.8/2.2/0.25,        % Inner right  
    -2/3.8/1.8/0.2,          % Outer left
    2/3.8/1.8/0.2,           % Outer right
    -0.6/3.8/2.0/0.22,       % Mid left
    0.6/3.8/2.0/0.22         % Mid right
} {
    % Golden cylinder body
    \filldraw[fill=gold!10, draw=gold!100!black, line width=1pt, sharp corners] 
        (\x-\r, \y) rectangle (\x+\r, \y-\h);
    
    % Connecting cable from top
    \draw[thick, gold!60, sharp corners] (\x, 4.2) -- (\x, \y);
    
    % Golden bands/ridges on cylinders
    \foreach \i in {0.3, 0.6, 0.9} {
        \draw[gold!90!black, line width=0.8pt, sharp corners] 
            (\x-\r, \y-\i) -- (\x+\r, \y-\i);
    }
}

% The quantum chip at the very bottom (center cylinder)
\filldraw[fill=gold!50, draw=gold!80!black, line width=2pt, sharp corners] 
    (-0.2, 1.3) rectangle (0.2, 1.0);
\end{tikzpicture}}};
    
     \node at (3.1,-6.05) {\scalebox{.25}{
        \begin{tikzpicture}[scale=1, every node/.style={inner sep=0, outer sep=0}]
        
        % Laptop base
        \fill[gray!70, sharp corners] (-3,-0.2) rectangle (3,0);
        \fill[gray!50, sharp corners] (-3,0) -- (3,0) -- (2.5,0.3) -- (-2.5,0.3) -- cycle;
        
        % Screen
        \fill[gray!40, sharp corners] (-2.5,0.3) rectangle (2.5,3);
        \fill[black!80!gray, sharp corners] (-2.4,0.4) rectangle (2.4,2.9); % screen content
        
        % Camera
        \fill[black!60, sharp corners] (0,2.85) circle (0.05);
        \end{tikzpicture}}};
    
\end{tikzpicture}}
    \caption{Proposed workflow to combine machine learning and quantum algorithms to obtain the universal functional.}
    \label{fig:WorkflowSimple}
\end{figure}

Specifically, our framework combines RDM-FT with deep neural networks (DNN) to learn functional data generated by variational quantum eigensolvers (VQE). The general workflow discussed in the following is summarized in Fig.~\ref{fig:WorkflowSimple}. A more detailed version of the workflow, including the specific methods employed in this publication, is given in Fig.~\ref{fig:WorkflowDetailed}. To ensure a well-learned functional, the DNN gets the data in the form of an intermediate functional which is more akin to a collection of density functionals with different off-diagonal terms of the single-body Hamiltonian $h$, which can be transformed easily into the full RDM-FT-Functional using the Hellmann-Feynman theorem \cite{Feynman1939, Schmidt1938}. We extend this approach through density matrix embedding theory (DMET) \cite{Knizia2013, Wouters2016, Mineh2022, Sekaran2023, Cernatic2024, Fromager2015, Sekaran2022}, developing a functional theoretic density matrix embedding theory (FT-DMET). This method achieves high accuracy in terms of the resulting energy, while building the fragment-system itself scales comparably to Hartree-Fock methods. The complexity of the calculations required for solving the smaller embedded problem, is fixed by the size of the fragment and independent of the total system size. Thus, our approach uses both quantum as well as classical methods in tandem to take advantage of quantum hardware with moderate qubit-counts \cite{Lerch2024}. However, depending on the quality of the preprocessing and the complexity of the system, larger embedded system can be required for good accuracies.

The paper is organized as follows. In Sec.~\ref{sec:TheoreticalBackground}, a review of RDM-FT's fundamentals, its connection to the aforementioned intermediate functional and the quantum algorithm are provided. Sec.~\ref{sec:ApplicationVQE} describes the adaptation of the quantum algorithm to RDM-FT, and the neural network implementation for learning the functional. Sec.~\ref{sec:DMET} demonstrates how combining the neural network approach with DMET (for fermionic and bosonic systems) enables broader applicability through fragment-based calculations. Finally, the methodology is exemplified on Bose-Hubbard and Fermi-Hubbard models of varying sizes as well as a two-band Fermi-Hubbard model in (Sec.~\ref{sec:Examples}).

%\newpage

\section{\label{sec:TheoreticalBackground}Fundamental Concepts and Methodologies}

This section provides an introduction into functional theories, with an emphasis on RDM-FT and the intermediate functional used in the following. Additionally, the specific quantum algorithm used here (VQE) is described in general terms.

\subsection{\label{subsec:RDM-FT}Reduced Density Matrix Functional Theory}

The many-body Hamiltonian investigated is of the general form
\begin{equation}\label{equ:QuantumChemistryHamiltonian}
  \hat{H} = \hat{T} + \hat{W} + \hat{V}_{\text{ext}},
\end{equation}
where $\hat{T}$, $\hat{W}$ and $\hat{V}_{\text{ext}}$ are the kinetic, the 2-body-interaction and the external potential contributions, respectively. These terms are expressed in second quantization. In general, any Hamiltonian that allows for a splitting into two or more parts can be described by a functional theory. This requires an additional quantity, like the density $\rho(r)$ or the one-particle reduced density matrix (1-RDM) $\gamma$ \cite{Xu2022}. The 1-RDM is the crucial quantity as it is the input variable of the RDM-functional and can be used for numerous different Hamiltonians, allowing for a higher degree of versatility compared to a density-functional. This gives more variability, especially within the context of lattice models like the Hubbard model, as we can freely change both the hopping strengths as well as the on-site external potentials. On the other hand, the DFT formalism for Hubbard models only accommodates changes in the on-site external potential. Additionally, within an orbital framework, the 1-RDM is a more natural choice when discussing changes in the potentials \cite{Giesbertz2019}. The spin-unpolarized 1-RDM is defined as
\begin{equation}
  \gamma_{ij} = \sum_{\sigma} \bra{\Psi} \left( \hat{a}_{\sigma, i}^\dagger \hat{a}_{\sigma, j} \right) \ket{\Psi},
\end{equation}
where the sum over $\sigma$ is over all spins, and the associated creation and annihilation operators, $\hat{a}_{\sigma, i}^{(\dagger)}$, act upon the many-body wavefunction $\ket{\Psi}$. A spin-dependent extension to the formalism is straightforward to define but is beyond the scope of this work. In Eq.~\eqref{equ:QuantumChemistryHamiltonian}, parts of the energy are then described by $\rho(r)$ and $\gamma$ as
\begin{align}\label{equ:SimpleEnergyContributions}
  E_{\text{ext}} &= \bra{\Psi} \hat{V}_{\text{ext}} \ket{\Psi} = \int dr \rho(r) v_{\text{ext}}(r), \\
  E_{\text{ext}} + E_{\text{kin}} &= \bra{\Psi} \hat{T} + \hat{V}_{\text{ext}} \ket{\Psi} = \text{Tr} \left( (T+v_{\text{ext}}) \gamma \right), \nonumber
\end{align}
for DFT and RDM-FT, respectively. Here, the hat-terms refer to the full many-body operators. $T$ and $v_{\text{ext}}$ are the coefficients of those operators, and are collected in a way such that a simple trace or integration with the density or trace with the reduced density matrix is sufficient to give their full energy contribution. The rest of the energy not contributing to Eq.~\eqref{equ:SimpleEnergyContributions} is collected in a mathematical object $\mathcal{F}$, which is referred to as the universal functional. For the non-degenerate case, a one-to-one relation between the ground state density, the ground state, the external potential and the functional $\mathcal{F}$, i.e., the Hohenberg-Kohn theorem \cite{Hohenberg1964}, is straightforward to show. This can be extended to observables being uniquely determined by the ground state density. Other quantities arising from other Hamiltonian-splittings also allow for such relations \cite{Gilbert1975, Wu2006, Xu2022}, giving the mathematical foundation for RDM-FT.

In order to find the universal functional $\mathcal{F}$ within RDM-FT variational principles can be used, with the Rayleigh-Ritz variational principle \cite{Ritz1909} being the most common. The ground state energy can be found by means of the following unconstrained minimization
\begin{equation}
 E_0(h=T+V_{\text{ext}}) = \min_{\ket{\Psi}} \left\{ \bra{\Psi} \hat{T} + \hat{W} + \hat{V}_{\text{ext}} \ket{\Psi} \right\} .
\end{equation}
Using the Functional-Theory inherent splitting of the Hamiltonian, a reduced-density-matrix-functional (in the following referred to as RDM-functional or $\mathcal{F}_{\text{RDM}}$) is defined as
\begin{equation}\label{equ:LevyLiebConstrainedSearch}
    \mathcal{F}_{\text{RDM}}[\gamma] = \min_{\ket{\Psi} \ \text{s.t.} \ \ket{\Psi}\mapsto \gamma} \left\{ \bra{\Psi} \hat{W} \ket{\Psi} \right\},
\end{equation}
which is Levy-Lieb's constrained search \cite{Levy1979}. In particular, the minimum is taken over all possible states that have a particular 1-RDM associated with them. As such, the ground state can be found with the much lower dimensional minimization
\begin{equation}
  E_0(h) = \min_{\gamma} \left\{ \mathcal{F}_{\text{RDM}}[\gamma] + \text{Tr}(h \gamma) \right\} .
\end{equation}
The relation between the ground state energy, as a functional of the single-body Hamiltonian $h$, and the RDM-functional $\mathcal{F}_{\text{RDM}}$ is known to be a Legendre-Fenchel transform \cite{Rockafellar1996, Schilling2018}, which ensures the ground state energy to be concave and the functional to be convex. Almost the entire complexity of solving the ground state problem has thus been shifted into obtaining the functional $\mathcal{F}_{\text{RDM}}$. Deriving the functional is of similar complexity as solving the ground state problem itself, and it has been shown to be a QMA-hard problem (Quantum-Merlin-Arthur) for the case of DFT \cite{Schuch2009}, with a proof that is applicable to any functional theory. Due to this fact, simply executing the constrained search on quantum hardware seems to be an inefficient way since directly solving the ground state problem is of similar complexity \cite{Wei2010}. However, for a sufficiently universal functional, as achieved here through DMET, the quantum computational effort gets shifted such that it needs to be done only once, and then the same functional can be reused repeatedly purely on classical hardware. 

The exact form of the functional is in general not known and using machine learning for this purpose shows great potential \cite{Koridon24}. The version of the functional which is learned in the following is directly inspired by RDM-FT. However, there are issues when learning the RDM-functional directly which will be discussed in more detail in Sec.~\ref{sec:ApplicationVQE}. To ease the learning procedure, the functional used in this work is a density functional where the off-diagonal elements of the single-body Hamiltonian are considered as additional variables
\begin{align}\label{equ:DNNFunctionalDefinition}
    \mathcal{F}_{\text{DNN}}&[n_0, ..., n_N; h_{01}, ..., h_{N-1 \ N}] \nonumber \\
    &= \min_{\ket{\Psi}\mapsto \boldsymbol{n}} \left\{ \bra{\Psi} \hat{W} + \sum_{ij, i\neq j} h_{ij} \hat{a}_i^\dagger \hat{a}_j \ket{\Psi} \right\} , \\ 
    &= E_0(h) - \sum_i n_i h_{ii} = E_0(h) - \sum_i \frac{d E_0(h)}{d h_{ii}} h_{ii} . \nonumber
\end{align}
In the following, this functional will be referred to as the Deep-Neural-Network- or DNN-functional $\mathcal{F}_{\text{DNN}}$. This approach, however, does not give direct access to the off-diagonal elements of the 1-RDM, which are required for the DMET scheme in order to avoid double counting energy contributions between the bath and the fragment. The simplest way around this - to make this DNN-functional effectively equivalent to regular RDM-FT - is to use Hellmann-Feynman's theorem \cite{Feynman1939, Schmidt1938} to connect the derivatives of the functionals as
\begin{align}\label{equ:HellmannFeynmanForFunctional}
    \frac{d}{d h_{ij}} \mathcal{F}_{\text{DNN}} =& \frac{d}{d h_{ij}} \left( E_0(h(n)) - \sum_k h_{kk} n_k \right) \nonumber ,\\
    =& \frac{d E_0(h)}{d h_{ij}} - \sum_k \frac{d h_{kk}}{d h_{ij}} n_k , \\
    =& \frac{\partial E_0}{\partial h_{ij}} + \sum_k \frac{\partial E_0}{\partial h_{kk}} \frac{d h_{kk}}{d h_{ij}} - \sum_k \frac{d h_{kk}}{d h_{ij}} n_k \nonumber , \\
    \stackrel{\makecell{\text{Hellmann-}\\ \text{Feynman}}}{=}& \gamma_{ij} + \sum_k n_k \frac{d h_{kk}}{d h_{ij}} - \sum_k \frac{d h_{kk}}{d h_{ij}} n_k \nonumber 
    = \gamma_{ij} \nonumber.
\end{align}
%directly with said off-diagonal 1-RDM element which will be needed later for the DMET-scheme.

The relation directly shows that the two functionals, $\mathcal{F}_{\text{RDM}}$ and $\mathcal{F}_{\text{DNN}}$, are effectively equivalent quantities, containing the same information but encoded slightly different. This is inspired by the relation between the partial derivative of the functional and the double occupancy in \cite{Senjean2018}. In terms of universality, this DNN-functional is also applicable to all single-body Hamiltonians, just like the RDM-functional, given they are inside the region covered by the training data. Thus, this redefinition is only to the benefit of the neural network.

\subsection{\label{subsec:VQE} Variational Quantum Eigensolver}
 
VQE was used as it has been demonstrated on NISQ hardware for quantum chemical systems \cite{
Google2020, Wang2024}. For large systems, particularly the Barren-plateau-issue makes executing the VQE-algorithm considerably less feasible \cite{Wang2021, Uvarov2021}. In such cases, other quantum algorithms, such as the QSE \cite{Yoshioka2022} or QPE \cite{OBrien2019}, can be slotted into the proposed workflow. 

For the VQE-algorithm, the wave function is parameterized with some vector $\theta$, i.e., ~$\ket{\Psi(\theta)}$. The parameter $\theta$ thereby directly affects the gates used on the level of the quantum hardware. This gives the slightly rewritten ground state problem
\begin{equation}\label{equ:GroundStateEnergyFromVQE}
    E_0 = \min_{\theta} \bra{\Psi(\theta)} \hat{H} \ket{\Psi(\theta)},
\end{equation}
where the minimization is over the parameter(s) $\theta$, and where $\hat{H}$ is the Hamiltonian expressed in Pauli-strings. The parameters $\theta$ are subsequently varied to minimize the energy expectation value \cite{AbuNada2021}. This minimization is done classically, while the evaluation of the expectation value is executed on a quantum processor. The algorithm is presented below as Alg.~\ref{alg:VQE}. Note that already for relatively small system sizes, the parameter $\theta$ can grow drastically in dimensionality if all possible states are to be represented. Since the introduction of VQE, efforts have been devoted to find good parametrizations that use physical and quantum chemical knowledge to reduce the size of the variational phase space by, e.g., utilizing coupled cluster like states \cite{Tilly2022} or complete active space approaches.

\vspace{0.5cm}

\begin{algorithm}[ht]
\caption{Variational Quantum Eigensolver}\label{alg:VQE}
\begin{algorithmic}[1]
\State fix starting parameters $\theta_{\text{initial}}$ \\
read-out the energy $\bra{\Psi(\theta)} \hat{H} \ket{\Psi(\theta)}$ \newline using QPUs \\
vary $\theta$ in accordance with a classical minimizer (e.g.~gradient descent) \\
repeat steps \small{2} and \small{3} until convergence \\
the approximate ground state energy \newline  is then $\bra{\Psi(\theta_{\text{min}})} \hat{H} \ket{\Psi(\theta_{\text{min}})}$ \\
additionally save resulting 1-RDM elements $\gamma_{ij} = \bra{\Psi (\theta_{\text{min}})} \hat{a}^\dagger_i \hat{a}_j \ket{\Psi (\theta_{\text{min}})}$ \newline read-out on QPUs
\end{algorithmic}
\end{algorithm}

\vspace{0.5cm}

As VQE is applied to both bosons and fermions, their mapping onto the quantum hardware needs to be considered carefully. Since qubits are essentially hard-core bosonic particles, the exchange symmetry does fit with the bosonic positive one. To allow for multiple occupations, a boson-to-qubit-mapping is required where now multiple qubits correspond to a single bosonic site at different occupation numbers \cite{Huang2021}. The most qubit-efficient  mapping scheme is then one utilizing a binary-type mapping where a state with a specific number of bosons in a specific site corresponds to the binary number of the qubits associated with this particular site ($\ket{6}\leftrightarrow\ket{...0110}$). Here, every basis state combination was encoded on one qubit due to the relatively small size for this proof of concept \footnote{ This is akin to one-hot encoding which while being unfavorable in terms of scaling was used in the following due to its simplicity. For the example of 2 bosons distributed across 2 sites the basis $\{\ket{2,0},\ket{1,1},\ket{0,2}\}$ then becomes $\{\ket{1,0,0},\ket{0,1,0},\ket{0,0,1}\}$ in terms of qubits with the appropriate interactions between the basis states being calculated classically. }. However, for larger settings the binary encoding would be required to ensure more favorable scaling. To enforce fermionic negative exchange symmetric nature on a quantum processors, well-known transformations like Jordan-Wigner- or Bravy-Kitaev-mappings  need to be employed \cite{Jordan1928, Bravyi2002, Tranter2018}. Specifically, the Jordan-Wigner transformation was used in obtaining the results presented in Sec.~\ref{subsec:FermionsExample}.

\subsection{\label{sec:ApplicationVQE} Finding the Functional with VQE}

% Having introduced the foundations of the formalism under investigation here, the specifics of obtaining the functional using VQE and neural networks are discussed in the following. The general formalism will be the same across the different particle types.

Applying VQE to the Levy-Lieb constrained search \eqref{equ:LevyLiebConstrainedSearch} results in a minimization with an adapted objective function
\begin{equation}\label{equ:ConstrainedSearchEquation}
    \mathcal{F}_{\text{RDM}}[\gamma] = \min_{\theta \ \text{s.t.} \ \ket{\Psi(\theta)}\mapsto\gamma} \bra{\Psi(\theta)} \hat{W} \ket{\Psi(\theta)} ,
\end{equation}
where the 1-RDM (or alternatively only the density) is fixed, which is a non-trivial task. The simplest way of dealing with the constrained search is to circumvent it. In the spirit of Refs.~\cite{Koridon24, Schade2022}, simple Lagrange multipliers are used to define a Hamiltonian
\begin{equation}\label{equ:GenericHamiltonian}
 \hat{H}(h) = \hat{W} + \sum_{ij} h_{ij} \hat{a}_i^\dagger \hat{a}_j ,
\end{equation}
scanning for different 1-RDMs. The single-body Hamiltonian terms, $h_{ij}$, take the role of the multipliers. After having minimized this particular Hamiltonian, one can then measure the expectation values of $\hat{a}^\dagger_i \hat{a}_j$ in order to access the 1-RDM of that particular ground state. Additionally, together with the ground state energy \eqref{equ:GroundStateEnergyFromVQE}, one can then calculate the RDM-functional value as
\begin{equation}\label{equ:RDMFunctionalCalculation}
 \mathcal{F}_{\text{RDM}}[\gamma] = E_0(h) - \sum_{ij} h_{ij} \gamma_{ij},
\end{equation}
by executing the Legendre transformation between the ground state energy and the functional directly. Since the RDM-functional is not learned directly but instead a version of the density functional which takes off-diagonal single-particle Hamiltonian-parts as additional input, this has to be slightly adjusted into the DNN-functional 
\begin{equation}\label{equ:InBetweenFunctionalCalculation}
 \mathcal{F}_{\text{DNN}}[n_0,...n_N; h_{01},...,h_{N-1 \ N}] = E_0(h) - \sum_{i} h_{ii} \gamma_{ii},
\end{equation}
as was introduced in Eq.~\ref{equ:DNNFunctionalDefinition}, where the diagonal elements of the 1RDM $\gamma_{ii}$ are equivalent to the densities $n_i$.

 %This is more favorable than directly learning the RDM-functional, \color{red} i.e., lacking training data due to non-differentiability and diverging gradients \color{black}.

\begin{figure}[ht]
\centering
\includegraphics[width=.95\linewidth]{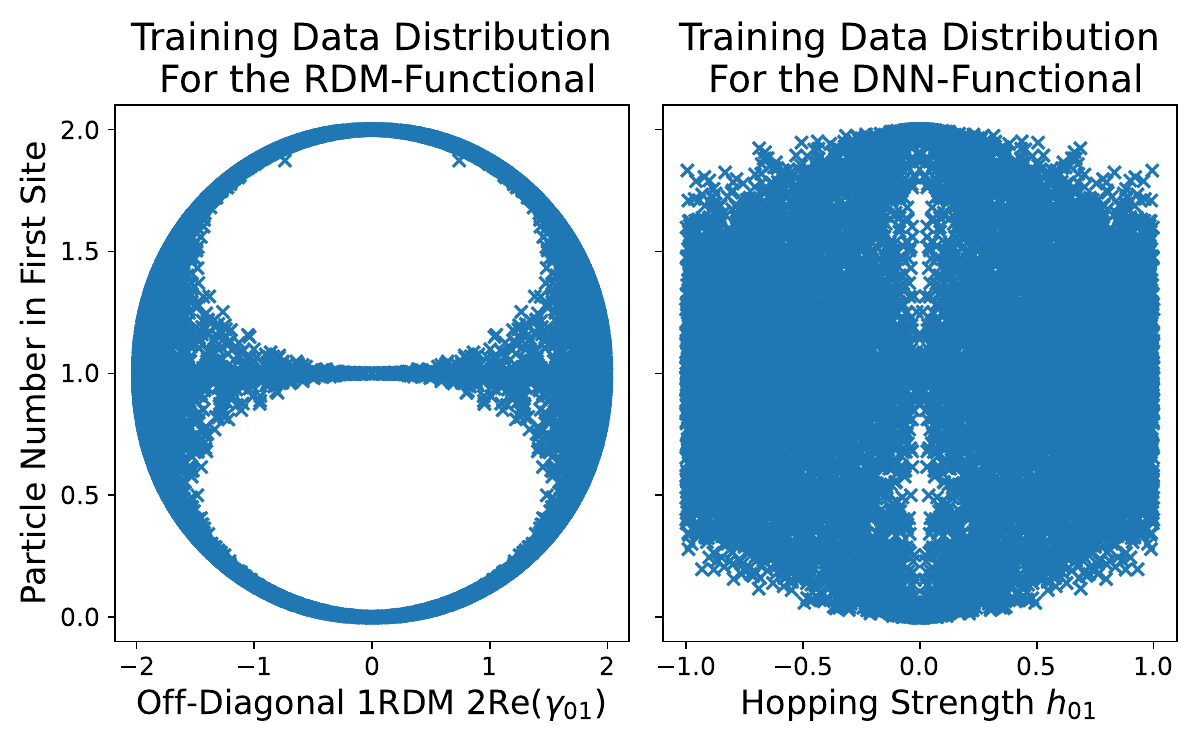}
\caption{\label{fig:TrainingDataDistribution} The distribution of the training data of the Fermi-Hubbard (Sec.~\ref{subsec:FermiHubbard}) model where the difference between RDM- and DNN-functional as the areas lacking training data for the RDM-case are filled for the DNN-training-data.}
\end{figure}

This procedure is then repeated for different single particle Hamiltonians $h$, resulting in different densities and also different $h_{ij}$. In other words, the ground state is found for different $h$-terms and then Legendre-transformed into the functional. To obtain the functional, we use a machine learning approach that requires both the densities as well as the off diagonal $h$-terms as inputs (see Fig.~\ref{fig:WorkflowSimple}). Non-differentiable points in the energy with respect to changes in the single-body Hamiltonian $h$ lead to a lack of training data for the RDM-functional (see Fig.~\ref{fig:TrainingDataDistribution}).
This is due to Eq.~\eqref{equ:RDMFunctionalCalculation} effectively being the Legendre-transformation of the ground state energy, leading to hyperplanes whenever the ground state energy exhibits a kink, with training data only at the ends of this hyperplane. Additionally, the RDM-functional exhibits diverging gradients towards the boundary of physically sensible 1-RDMs \cite{Schilling2019}, increasing the difficulty for a RELU-based neural network to represent the functional values close to this boundary. Therefore, it is easier to learn the DNN-functional as defined in Eq.~\ref{equ:InBetweenFunctionalCalculation}. This DNN-functional is a Legendre transformation of the full RDM-functional with respect to the off-diagonal 1-RDM elements, which leads to the diverging gradients becoming linear for large off-diagonal single-body Hamiltonian terms. Nevertheless, the requirement of a gradient in the following equations does require a high accuracy in the neural network. Specifically, we use deep neural networks (DNNs), which allows us to reduce the number of such required values by recognising patterns in the data that functionals are expected to have \cite{Shao2023, Schmidt2021}. A fully connected multi-layer neural network using the RELU activation function was chosen. For larger two-dimensional systems, a convolutional layer emphasizing local influences of densities on each other could be beneficial. However, such specialized layers are not needed for the purpose of the current paper. In principle, provided with sufficient training data, with a noiseless QPU and with a sufficiently complex neural network, an accuracy of the energy comparable to a full CI calculation would be possible. 

Our strategy allows the functional to be accurate in regimes where regular DFT tends to struggle. Importantly, minor changes in the 1-RDM lead only to minor changes in the wave-function and thereby the functional value as long as there is no level-crossing and the ground state energy being strictly concave. A more rigorous investigation of this consideration is given in appendix \ref{appendix:SimilarDensitiesSimilarWavefunctions} and \cite{DAmico2011}. Minor errors in the quantum processors should, therefore, only lead to minor inaccuracies in the machine learning model, as can be seen by the results in Sec.~\ref{sec:Examples}.

The proposed method has the benefit that every run counts towards the final functional, meaning that every run is used for training the DNN. This is a natural advantage over methods which need to adjust their Lagrange multipliers over multiple runs.
Furthermore, the scheme can be heavily parallelized, since the different runs with different single-body Hamiltonians are fully independent from one-another.
Additionally, the topic of v-representability is not an issue in our approach. A 1-RDM (or a density) is said to be v-representable if and only if it is the ground state 1-RDM (or the ground state density) to a single-body Hamiltonian i.e., can be derived from a ground state \cite{Liebert2023}. Since every contribution to the training data set is by its very definition the ground state 1-RDM to a specific Hamiltonian, every data point is v-representable\footnote{The resulting functional is the lower convex envelop of the pure state functional. It corresponds to the ensemble state functional - a fact that however does not influence the energies and all other quantities derived from the functional.}. Now, while the training data is by its very definition v-representable, the neural network in principle would give data for non-representable input as well meaning that it should be ensured the densities inserted are physical in the sense that they obey the Pauli-exclusion principle and add up to the correct particle number.

The VQE-scheme chosen here is the variational Hamiltonian Ansatz as it is a tried and tested approach \cite{Wecker2015,Wecker20152}, which allows for relatively short circuits. In it, the quantum state gets varied in accordance with the given Hamiltonian i.e.
\begin{equation}
 \ket{\Psi(\theta_j^k)} = \Pi_j e^{-i \hat{T} \theta_j^0} e^{-i \hat{V}_{\text{ext}} \theta_j^1} e^{-i \hat{W} \theta_j^2} \ket{\Psi_0} .
\end{equation}
This promises both efficiency and accuracy as it is inspired by a trotterized time evolution.

\section{\label{sec:DMET}Density Matrix Embedding Theory with Functional Theory}

Finding the exact functional takes considerably more resources than directly solving the ground state problem itself. Functional theory, however, thrives by exploiting its increased universality compared to simply working out one singular ground state problem itself. The RDM-functional, and by extension the DNN-functional defined in Eq.~\eqref{equ:InBetweenFunctionalCalculation}, is particularly useful as it is only defined by the interactions in the system under study. This means that once obtained, the same functional can be reused for any system where the interactions are identical, independently of the kinetic or external potential part. To fully exploit this fact, even for systems that are considerably larger and more complex than the system on which the functional was derived, DMET can be used~\cite{Knizia2013, Wouters2016, Cernatic2024, Sekaran2023}. This section focuses on integrating the functional theory into a regular DMET workflow.

\subsection{DMET: A General Introduction}

At its core, DMET partitions a large system into a fragment and an environment, and the effect of the environment on the fragment is then described by a much smaller bath. In general, a quantum state can be partitioned into two subsystems, the environment with states associated with it denoted as $\ket{E_i}$ and the fragment's states $\ket{F_i}$, resulting in a full description as $\ket{\Psi} = \sum_{ij} c_{ij} \ket{F_i} \ket{E_j}$ where the number of addends is arbitrarily large. By transforming into a different basis set using a singular value decomposition (SVD), this number of required terms can be greatly reduced \cite{Wouters2016, Schollwck2011}. In particular, the smaller of the two partitions gives an upper limit for the number of required number of terms in the sum
\begin{align}
 \ket{\Psi} &= \sum_{ij} c_{ij} \ket{F_i} \ket{E_j} \stackrel{\text{SVD}}{=} \sum_{ijk} \lambda_k U_{ik} \ket{F_i} V_{jk} \ket{E_k} \nonumber , \\
 &= \sum_{k=0}^{\min\{ \text{dim}(F), \text{dim}(B) \}} \lambda_k \ket{\Tilde{F}_k} \ket{\Tilde{E}_k} , 
\end{align}
where the unitary $U$ is derived from the singular value decomposition, and the new states $\ket{\Tilde{F}_k}$ and $\ket{\Tilde{E}_k}$ are the previous fragment and environment states now transformed into the new basis. This limit of required elements in the sum is independent of the size and complexity of the environment. The environment states can then be used to define a space of bath orbitals in order to gauge the effects of the environment on the fragment. Solving the general system using a mean-field method (MF) allows for the construction of an approximate Schmidt decomposition, while keeping the computational effort limited. In particular, a projection $P$ onto this smaller space can thus be defined (e.g.~see Eq.~\ref{equ:fermionicProjector} explicitly). Dependent on the quality of the mean-field-preprocessing, as well as the inherent complexity of the system under investigation, larger fragment-bath-systems can be needed to achieve a good accuracy in terms of the final energy result. The details of constructing the projectors are discussed in Sec.~\ref{subsec:BosonicFTDMET} and \ref{subsec:FermionicFTDMET}, as they differ slightly for fermionic and bosonic systems. The general Hamiltonian becomes
\begin{align}\label{equ:CreateEmbeddingHamiltonian}
 \hat{H} &= \hat{W} + \hat{h} \rightarrow \hat{H}_{\text{emb}} = \hat{W}^{\text{emb}} + \hat{h}^{\text{emb}}  \nonumber  \\
 &= P^\dagger P^\dagger \hat{W} P P + P^\dagger \hat{h} P ,
\end{align}
acting only on the fragment-bath subsystem. This is in the following referred to as embedded system. In this equation, the non sub-/superscripted operators act on the whole system, while the embedded operators act only on the fragment-bath system, with P being the projector between those two systems. Solving this with a more accurate method self-consistently allows for a better energy estimation for the full system compared to the mean-field result \cite{Knizia2013, Wouters2016}. Additionally,  insights into observables like the density can be gained in this way \cite{Mineh2022}.

While solving $\hat{H}_{\text{emb}}$ is a straightforward task, a self-consistency condition needs to be imposed to ensure that the number of particles in the fragment is consistent with the rest of the system. This is conventionally done by imposing a uniform chemical potential on the fragment, which is then tuned until the target is sufficiently met, requiring multiple runs of the more exact solver on the embedded system. Additional conditions, like off-diagonal elements of the 1-RDM, can also be enforced \cite{Wouters2016, Mineh2022}. The entire workflow of the here described version of one-shot DMET is summarized in Alg.~\ref{alg:DMET}.

\vspace{0.5cm}

\begin{algorithm}[ht]
\caption{Density Matrix Embedding \newline Theory (with non-interacting baths)}\label{alg:DMET}
\begin{algorithmic}[1]
\State solve the large system with a mean-field method and obtain the approximate ground state $\ket{\Psi_{\text{MF}}}$ \\
construct the mean-field ground state 1-RDM $\gamma_{\text{MF}}$ \\
diagonalize $\gamma_{\text{MF}}^{\text{env}}$ where the entries associated with the fragment are removed \\
construct a projector $P$ as the collection of the eigenvalues from step \small{3} \\
project onto the embedded system $\hat{H}^{\text{emb}} = P^\dagger \hat{h} P + \hat{W}^{\text{frag}}$ \\
solve the embedded problem where self-consistency needs to be enforced (e.g.~by adding an appropriate chemical potential on the fragment) \\
the resulting total energy is then $E_0 = \sum_{i \in \text{frag}} \left( E_0^{\text{emb}} - \frac{1}{2} h^{\text{frag-bath}} \gamma^{\text{frag-bath}} \right)$
\end{algorithmic}
\end{algorithm}

\vspace{0.5cm}

\newpage

\subsection{Functional Theoretic DMET}

The RDM-functional can be slotted into the workflow of regular DMET to solve the embedded Hamiltonian. The interacting term in the fragment-bath system can be chosen to be either fully interacting, i.e.~the full $P^\dagger P^\dagger \hat{W} P P$, or only interacting on the fragment, i.e.~a non-interacting bath where the interaction only affects the fragment itself \cite{Mineh2022}. Although the latter is in general less accurate, it does converge to the exact result when increasing the fragment size \cite{Wouters2016}. Here, the non-interacting formalism is used, as it allows for the RDM-functional to be reused whenever the fragment's interaction is identical. Therefore, the functional can be reused indifferent of any changes in geometry, hopping terms or any other parameters in the initial system, as long as the prior condition is met. For regular RDM-FT, the equation that needs to be minimized then becomes
\begin{align}\label{equ:FTDMETMinimization}
 E_0^{\text{emb}} = \min_{\gamma} \bigg\{ &\mathcal{F}_{\text{RDM}}[\gamma] \ \\
 &+ \sum_{i,j\in\text{emb}} h_{ij}^{\text{emb}} \gamma_{ij} \bigg\} ,  \nonumber
\end{align}
in which the functional $\mathcal{F}$ accounts only for the interaction within the fragment. For our purposes, we use the DNN-functional as defined in Eq.~\eqref{equ:InBetweenFunctionalCalculation} leading to the following minimization
\begin{align}\label{equ:FTDMETMinimizationWithDNN}
    E_0^{\text{emb}} = \min_{n} \bigg\{ &\mathcal{F}_{\text{DNN}}[n_0,...; h^{\text{emb}}_{01}, ...] \\
    &+ \sum_i h^{\text{emb}}_{ii} n_i \bigg\}. \nonumber
\end{align}
Here the DNN-functional is equivalent to a density functional defined with the appropriate hopping terms of the embedded single-body Hamiltonian.

%An approximate reintroduction of interactions between fragment and bath is made in section \ref{subsec:FermionicFTDMET}. 
One advantage of this approach over the regular one-shot DMET, is the direct enforcement of self-consistency constraints. Since the particle numbers are directly tunable as input variables into the machine-learned functional, such restrictions can be directly imposed. Therefore, no additional self-consistency-runs to optimize the chemical potential are required.

The result of Eq.~\eqref{equ:FTDMETMinimization} does not directly give the ground state energy. Instead, it finds the minimizing 1-RDM
\begin{equation}
   \bar{\gamma}^{\text{emb}} = \argmin_{\gamma^{\text{emb}}} \left\{ \mathcal{F}_{\text{RDM}}[\gamma^{\text{emb}}] + \sum_{i,j\in\text{emb}} h_{ij}^{\text{emb}} \gamma_{ij}^{\text{emb}} \right\}, 
\end{equation}
where the bar indicates the minimizing argument. To get access to the same quantity from the DNN-functional, the Hellmann-Feynman theorem is employed (see Eq.~\ref{equ:HellmannFeynmanForFunctional}) i.e.
\begin{multline}
    \bar{\gamma}^{\text{emb}}_{ij} = \frac{1}{2\delta} \Big(\mathcal{F}_{\text{DNN}}[\bar{n}_0, ...; ..., h^{\text{emb}}_{ij} + \delta, ...] \\
    - \mathcal{F}_{\text{DNN}}[\bar{n}_0, ...; ..., h^{\text{emb}}_{ij} - \delta, ...]\Big),
\end{multline}
where $\bar{n}_i$ are the densities found from the previous minimization. This scheme has the added advantage over the regular RDM-FT that the variational space in the minimization is further reduced, as only the densities are variables which need to be changed. This can be used to train the neural network around the area of interest only in terms of the hopping strengths, which can increase accuracy and reduces the amount of required training data at the cost of universality.

Subsequently, this $\bar{\gamma}$ is inserted into the equation
\begin{multline}\label{equ:FTDMETEnergy}
 E_0^{\text{frag}} = \mathcal{F}_{\text{DNN}}[\bar{n}_0,...; h^{\text{emb}}_{01}, ...] + \sum_{i\in\text{frag}} h^{\text{emb}}_{ii} \bar{n}_i \\
 - \sum_{i,j\in\text{bath}} h_{ij}^{\text{emb}} \bar{\gamma}_{ij}^{\text{emb}}  \\
 - \frac{1}{2} \bigg( \sum_{i\in\text{frag},j\in\text{bath}} h_{ij}^{\text{emb}} \bar{\gamma}_{ij}^{\text{emb}} \hspace{1.5cm} \\ 
 + \sum_{i\in\text{bath},j\in\text{frag}} h_{ij}^{\text{emb}} \bar{\gamma}_{ij}^{\text{emb}}  \bigg) , 
\end{multline}
where the factor of $\frac{1}{2}$ avoids double counting the energy contributions of hopping terms between different fragments. This is unique to the used definition of $\mathcal{F}_{\text{DNN}}$, where the energy contribution is already included in the quantity of the functional itself and thus needs to be subtracted for the bath sites. If the system is made up of identical fragments, this energy has to be multiplied by the fragment number. For a heterogeneous model the calculation needs to be repeated for each unique fragment. This can entail additional optimization loops if the initial guess of the distribution of particles between the fragments is optimized as well and not kept at the initial guess. This is not to be confused with the fragment-bath particle distribution, which is fixed by the functional formalism.

As long as observables can be expressed purely in the form of single-body terms, or can be directly derived from the functional, the minimizing 1-RDM $\bar{\gamma}$ also allows for insights into observables. Similar to the procedure presented before, the observable $\hat{O}$ is projected down onto the embedded system $\hat{O}^{\text{emb}} = P^\dagger \hat{O} P$ to be used in conjunction with $\bar{\gamma}$ to get $\langle \hat{O}^{\text{frag}} \rangle = \text{Tr}(O^{\text{emb}} \bar{\gamma})$, where again the pure bath contributions need to be discarded, while the fragment-bath contributions need to be halved to avoid double counting.

\subsection{\label{subsec:BosonicFTDMET} Bosonic FT-DMET}

DMET used on bosonic systems is an application that so far has only been subject to limited study \cite{Sandhoefer2016}. Since bosons do not abide to an exclusion principle like fermions, a Bose-Einstein condensation of all particles into a singular state is possible. The computationally cheap initial calculation is done here by diagonalizing the single-body Hamiltonian $h$, an operation with the associated computational complexity of $\mathcal{O}(N^3)$, a QR-decomposition \cite{Golub2013}. Its eigenvectors build an approximate 1-RDM of the complete system, with the lowest energy eigenvalue being exclusively used to obtain
\begin{equation}
 \gamma_{\text{MF}} = (v_0, v_0, v_0, ...) \cdot (v_0, v_0, v_0, ...)^T,
\end{equation}
where $v_0$ is the eigenvector corresponding to the lowest eigenvalue and it is used $N_{b}$-times, with $N_{b}$ being the number of bosonic particles in the entire system. This effectively gives a Bose-Einstein condensate, as no interactions are yet considered that penalize such behaviour.

The environment's 1-RDM is then just the mean-field 1-RDM $\gamma_{MF}$ with the fragment's orbitals being removed. The $(N-N_{\text{frag}})\times (N-N_{\text{frag}})$-matrix, with $N_{\text{frag}}$ the number of fragment orbitals, is diagonalized in the next step (scaling as $\mathcal{O}((N-N_{\text{frag}})^3)$ if again a QR-decomposition is used \cite{Golub2013}). All its eigenvalues will be $0$ with a singular exception which has the eigenvalue $n_{env}$, the number of particles in the whole environment. See the appendix \ref{appendix:rank1matrix} for the more rigorous argument. The particular eigenvector $v_0$ corresponding to the non-zero eigenstate defines the projection
\begin{equation}\label{equ:BosonicProjector}
 P_{\text{Boson}} = \begin{pmatrix} \mathds{1}_{N_{\text{frag}}} & 0 \\ 0 & v_0 \end{pmatrix} ,
\end{equation}
onto a fragment-bath system which can thus easily be solved using the RDM-functional. The construction of the embedding system from this projector then scales like $\mathcal{O}(N^2 N_{\text{emb}})$, where $N_{\text{emb}}$ is the number of sites in the embedded system \cite{Golub2013}. A consequence of this particular structure is that the projector is a $(N_{\text{frag}}+1)\times N$-matrix, meaning that only a singular bath site is required to execute this particular scheme. More sophisticated approximations, which do not lead to complete Bose-Einstein condensation on the mean-field level, would however require more bath sites. The embedded system still has the same number of bosons as the initial system, meaning that studying a larger-sized system with a constant particle number density increases the complexity of the fragment-bath system. Note that the complexity is still drastically reduced compared to the initial exact problem. This is exemplified with the example of the Bose-Hubbard model in Sec.~\ref{subsec:BoseHubbard}. Since solving the minimization in Eq.~\eqref{equ:FTDMETMinimization} does not scale with the system size, the total computational complexity of this bosonic FT-DMET algorithm scales like $\mathcal{O}(N^3)$. This is a strong reduction with respect to exact diagonalization, which scales exponentially due to the exponential dimensionality of the Hilbert space.

\subsection{\label{subsec:FermionicFTDMET} Fermionic FT-DMET}

In the case of fermions, we first assume that the interactions between fragment and bath are negligible, which is inline with the non-interacting bath formalism \cite{Wouters2016, Mineh2022}. In this case, one can once again employ the scheme of solving the entire Hamiltonian on a mean-field level, diagonalising the single-body Hamiltonian and use this to build an approximate 1-RDM $\gamma_{MF}$. Its complexity scaling is again $\mathcal{O}(N^3)$, as the most expensive step is diagonalizing the matrix $h$, which has a number of rows and columns equal to the number of sites or orbitals \footnote{In more general quantum chemical settings, this would take a Hartree-Fock calculation scaling in theory like $\mathcal{O}(N^4)$, but effectively scales a little bit worse than $N^2$ \cite{Strout1995}. Building the embedded system contributes also with $\mathcal{O}(N^3)$ for the diagonalisation and other transformations. The transformation of the 2-electron-repulsion-integrals into orthogonalised orbitals have a higher cost at $\mathcal{O}(N^5)$ exploiting symmetries and sparsity \protect\cite{Simons2023}. Comparing this to other quantum-chemical methods, this is a clear improvement over the exponential scaling of FCI, as well as having the potential to being an improvement over coupled-cluster methods (at worst $\mathcal{O}(N^6)$ for CCSD, $\mathcal{O}(N^7)$ for CCSD(T) and $\mathcal{O}(N^8)$ for CCSDT \protect\cite{Crawford2007}) given that an appropriate reusable fragment and 2-electron repulsion is chosen. The size of the fragment necessary to gauge the important energy contributions is however system-dependent.}. The next step is again to remove the fragment's rows and columns and diagonalizing the rump 1-RDM. This results in three different types of eigenvalues. They are either 0, 1 (or 2 accounting for electronic spins), or a value in-between. The number of sites with in-between eigenvalues is equal to the number of fragment sites \cite{Mineh2022}. These particular values are crucial for building up the fragment using a projector
\begin{equation}\label{equ:fermionicProjector}
 P_{\text{Fermion}} = \begin{pmatrix} \mathds{1}_{N_{\text{frag}}} & 0 & 0 & ...\\ 0 & v_0 & v_1 & ... \end{pmatrix},
\end{equation}
where the eigenvectors $v_i$ correspond to these in-between eigenvalues ($0 < \lambda_i < 1$). The resulting ground state problem becomes
\begin{equation}
 H_{\text{emb}} = h^{\text{emb}} + W^{\text{frag}} = P_{\text{Fermion}}^\dagger h P_{\text{Fermion}} + W^{\text{frag}},
\end{equation}
in the non-interacting formalism. For systems with strong on-site interactions and negligible interactions between fragments, this formulation is sufficient as shown for the Fermi-Hubbard model (Sec.~\ref{subsec:FermiHubbard}).

\section{\label{sec:Examples} Results}

The scheme of using quantum computing to find the functional and to enlarge its universality with the DMET-formalism is executed for a few test cases in this section. For bosons, the Bose-Hubbard model is analysed in Sec.~\ref{subsec:BoseHubbard}. The formalism for fermions is on display in Sec.~\ref{subsec:FermionsExample}, first for the single-band Fermi-Hubbard model, while a two-band Fermi-Hubbard model is used as an example for more complex systems requiring a 8-qubit QPU for generating the learning data.

\subsection{\label{subsec:BoseHubbard} Bose-Hubbard Model}

In its general form, the Bose-Hubbard Hamiltonian is given by
\begin{align}
 H(v_{\text{ext}},t,w) = &W + T + v_{\text{ext}} \nonumber , \\
 = &w \sum_{i=0}^N \left( \hat{a}^\dagger_i \hat{a}_i \hat{a}^\dagger_i \hat{a}_i - 1 \right) \\
 &+ \sum_{i,j=0}^N t_{ij} \hat{a}^\dagger_i \hat{a}_j + \sum_{i=0}^N v_{\text{ext}}^{(i)} \hat{a}_i^\dagger \hat{a}_i \nonumber ,
\end{align}
for $N$ lattice sites \cite{Fisher1989, Freericks1994, Elstner1999, SafaviNaini2012}. 
In general, an external potential $v_{\text{ext}}^{(i)}$ can be added onto the regular formalism. Additionally, different hopping strengths would also be possible without any additional calculations on the quantum processor for the FT-DMET formalism, as it is applicable to any combination of single-body Hamiltonian terms $h$. To keep this test-case concise and to exploit translational invariance, the hopping term as well as the external potential are assumed to be uniform, with hopping only between nearest neighbours. In cases lacking this invariance, multiple fragments need to be calculated using the same functional. Thus, the complexity would only increase linearly in the number of fragments, but these calculations can be executed in parallel.

% as due to the lack of translational symmetry, the result cannot just be multiplied with the number of fragments in that case.

\begin{figure}[ht]
\centering
\includegraphics[width=.95\linewidth]{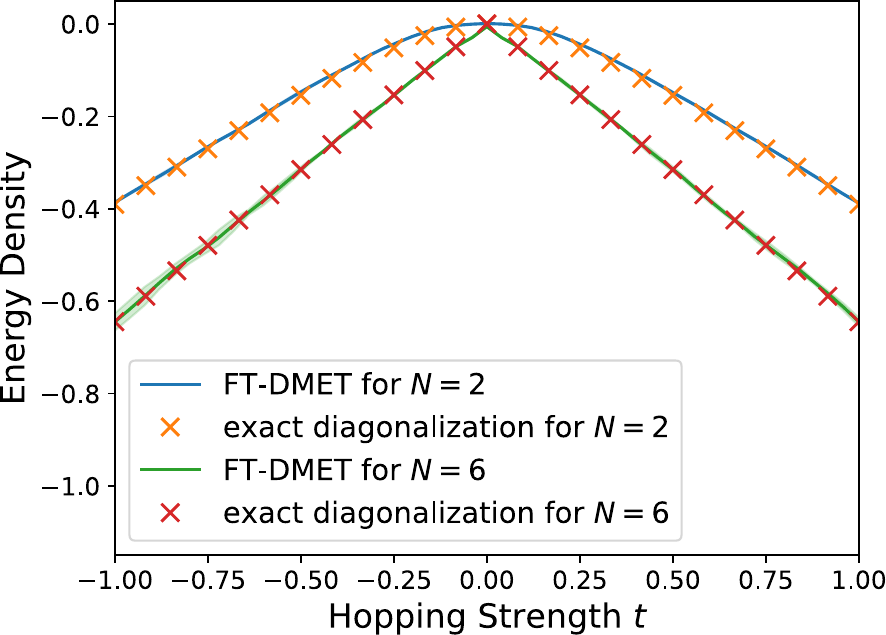}
\caption{\label{fig:BoseHubbardDMET} The energy of a Bose-Hubbard chain with 2 bosons at a length of 2 and 6 sites with a changing hopping term, while the interaction strength remains fixed to 1. The standard deviation (shaded) was found by training the DNN-functional five times on the same data.}
\end{figure}

\begin{figure}[ht]
\begin{center}
\includegraphics[width=.95\linewidth]{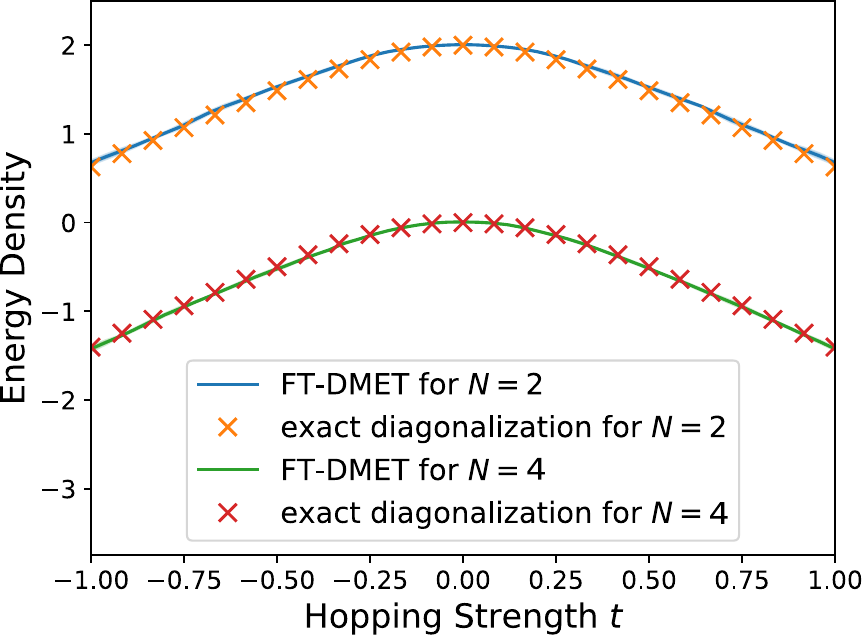}
\end{center}
\caption{\label{fig:BoseHubbardDMETN4} The Bose-Hubbard chain now with 4 bosons at a length of 2 and 4 sites. The standard deviation (shaded) was found by training the DNN-functional five times on the same data.}
\end{figure}

As a proof of concept, we use a Hubbard dimer with two bosons to derive the functional. One site, in agreement with the DMET-formalism, was kept as non-interacting. The on-site interaction on the 0-th site is set to 1 while the other site is completely non-interacting. Throughout this work, the quantum computing side of the code was implemented using Cirq (version 1.3.0 \cite{CirqDevelopers_2024}) and OpenFermion (Version 1.6.1 \cite{McClean2020}). Due to the uniform nature of the Bose-Hubbard-interaction, the functional for any other interaction strength can be derived directly from those results. This holds as long as the sign does not switch, as in such case the factor cannot be drawn out of the minimization in Eq.~\eqref{equ:ConstrainedSearchEquation}. One can directly read of the expectation value of the hopping terms i.e.~the 1-RDM elements, as the slope of the DNN-functional (in accordance with Eq.~\ref{equ:HellmannFeynmanForFunctional}). Here the slope is equivalent to $2\text{Re}(\gamma_{01})$, due to the expectation value being $\langle \hat{a}_0^\dagger \hat{a}_1 + \hat{a}_1^\dagger \hat{a}_0 \rangle$. This not only already contains all the information required for our purposes, but has the added advantage over $\gamma_{01}=\langle \hat{a}_0^\dagger \hat{a}_1 \rangle$ of being self-adjoint thus being more straight forward to evaluate. The results obtained by using this DNN-functional in the context of DMET, described in Sec.~\ref{subsec:BosonicFTDMET}, are shown in Fig.~\ref{fig:BoseHubbardDMET} and \ref{fig:BoseHubbardDMETN4}.

Both examples studied here describe a periodic, one-dimensional chain. Since the entirety of the bosonic particles are condensed into the bath site apart from the particles in the fragment, the DNN-functional derived above for this model restricts us to only 2 bosons in Fig.~\ref{fig:BoseHubbardDMET}. To show the robustness of this formalism, five DNN-functionals where trained on the same training data set (see the shaded area for the standard deviation between the functionals). This was extended to 4 bosons, see Fig.~\ref{fig:BoseHubbardDMETN4}, where the DNN-functionals trained for 4 particles shows very good agreement amongst themselves as well as with the exact result. The Bose-Hubbard chains were chosen to be only of limited size, as larger chain lengths effectively lead to linear energies. Such is the case for $N=6$, see Fig.~\ref{fig:BoseHubbardDMET}, for which the chain is so sparsely populated that the on-site interaction does not really affect their behaviour. It does, however, in general not impose any restrictions on lattice size or geometry.

\subsection{\label{subsec:FermionsExample} Fermions}

For the results of the single- and two-band Fermi-Hubbard model in the following section, the fermionic FT-DMET formalism described in Sec.~\ref{subsec:FermionicFTDMET} is utilized.

\subsubsection{\label{subsec:FermiHubbard} Fermions: Single-Band Fermi-Hubbard Model}

Despite its simple mathematical form, the Fermi-Hubbard model is an ideal test case for highly-correlated systems \cite{Hubbard1963, Edit2013, Cade2020}. The Hamiltonian consists of three competing terms
\begin{align}
 H(v_{\text{ext}}, t, w) = &w \sum_{i=0}^{N-1} \hat{a}^\dagger_{i\uparrow} \hat{a}^\dagger_{i\downarrow} \hat{a}_{i\downarrow} \hat{a}_{i\uparrow} \\
  &+ \sum_{i,j=0, \sigma\in\{\uparrow,\downarrow\} }^{N-1} t_{ij} \hat{a}^\dagger_{i\sigma} \hat{a}^\dagger_{j\sigma} \nonumber \\
 &+ \sum_{i=0, \sigma\in\{\uparrow,\downarrow\} }^{N-1} v_{\text{ext}}^{(i)} \hat{a}^\dagger_{i\sigma} \hat{a}_{i\sigma}, \nonumber
\end{align}
similar to the prior discussed Bose-Hubbard model. The main difference is the odd spin of the particles, in particular either up or down spins and, as a consequence of this, that the highest occupation number of a single site is restricted to two due to the Pauli exclusion principle.

% paired with their wave-functions being negative under exchange. Daniel: eliminated this sentence as it is equivalent to the Pauli principle.

\begin{figure}[ht]
\centering
\includegraphics[width=.95\linewidth]{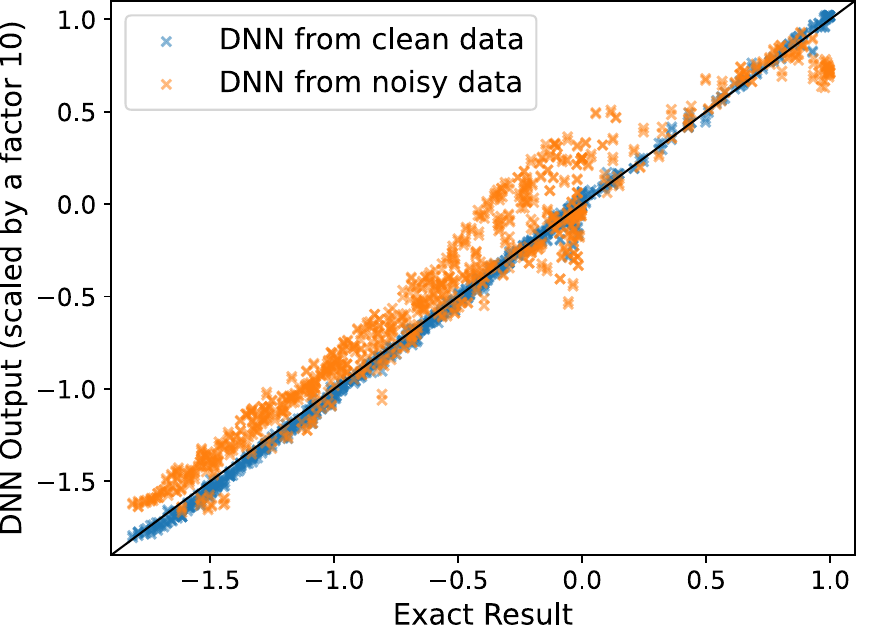}
\caption{\label{fig:MLQualityFermionicRDMFunctional} The exact functional value $\mathcal{F}_{\text{DNN}}$ (Eq.~\ref{equ:InBetweenFunctionalCalculation}) is plotted against the output of the machine learned functional with it being trained on noiseless (blue) and noisy (orange) quantum to test its accuracy on random inputs. The error in y-direction is increased by a factor of 10 to increase readability.}
\end{figure}

\begin{figure}[ht]
\centering
\includegraphics[width=.95\linewidth]{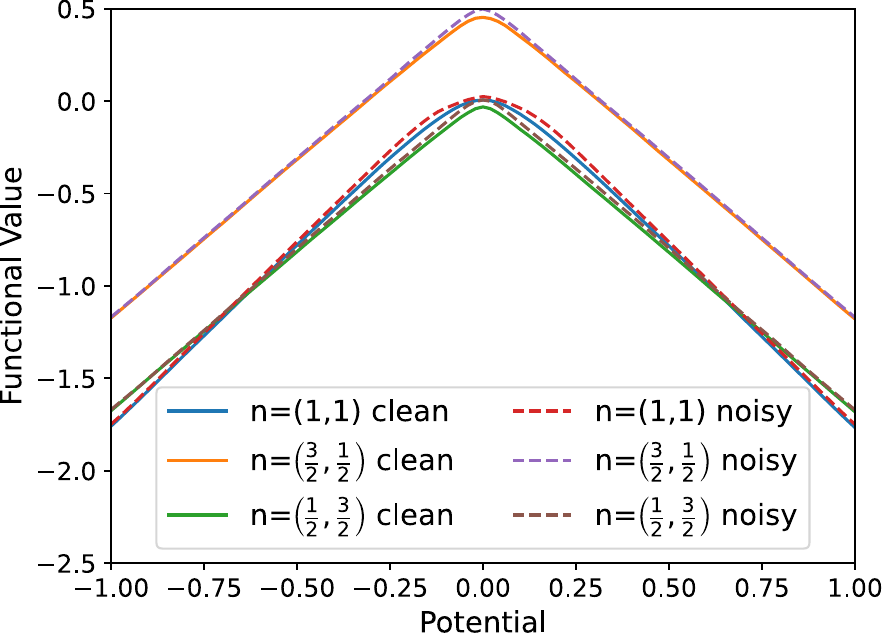}
\caption{\label{fig:FermionicRDMFunctional} The DNN-functional for the Fermi-Hubbard model with a total particle number of 2 fermions from an exact (solid) and noisy (dashed) quantum processor. The on-site interaction on the 0-th site is set to 1 while the other, bath-related, site is completely non-interacting. The labels here refer to the particle densities in each site with the first value referring to the density in the fragment and the second to the density in the bath site.}
\end{figure}

The functional is again derived for a minimal model of two lattice sites, resulting in four qubits. In order to assess the quality of the machine-learned functional, a comparison between our results and the exact functional is presented in Fig.~\ref{fig:MLQualityFermionicRDMFunctional}. To obtain the exact result, a state-vector was simulated propagating through the VQE quantum circuit. The noisy model used Cirq's Sycamore simulator (version 1.3.0 \cite{CirqDevelopers_2024}) with a shot number of 5.000 using the Weber processor and its provided median noise. To limit the resulting errors, the state was read-out only in the Z-basis, from which the energy can be directly calculated. Additionally, we used the minimal result of the entire VQE-process, which due to the noise did not necessarily coincide with the final state of the VQE-minimization. This can be seen as a simple form of error mitigation possible for this particular model. In general, reading out additional other Pauli strings will lead to additional noise in the data, leading to worse results down the line. In particular, the functional's value is shifted upwards systematically due to the noisy VQE struggling to find the exact ground state \cite{Koridon24}. These errors seem to be only minor when looking at the DNN-functional directly, see Fig.~\ref{fig:FermionicRDMFunctional}, but due to this formalism requiring the derivative, already minor errors can lead to issues. To truly take advantage of the proposed formalism, error mitigation \cite{GiurgicaTiron2020, Maciejewski2020} is a necessity and compatible with the workflow presented here (see Fig.~\ref{fig:WorkflowDetailed}).

The on-site interaction on the fragment-related site is set to $+1$, meaning it is compatible with any positive interaction strength as, due to linearity, any positive factor can be taken out of the minimization in Eq.~\eqref{equ:ConstrainedSearchEquation}. For a negative interaction strength, however, a new functional resulting from a new set of training data would be required.

\begin{figure}
\includegraphics[width=.95\linewidth]{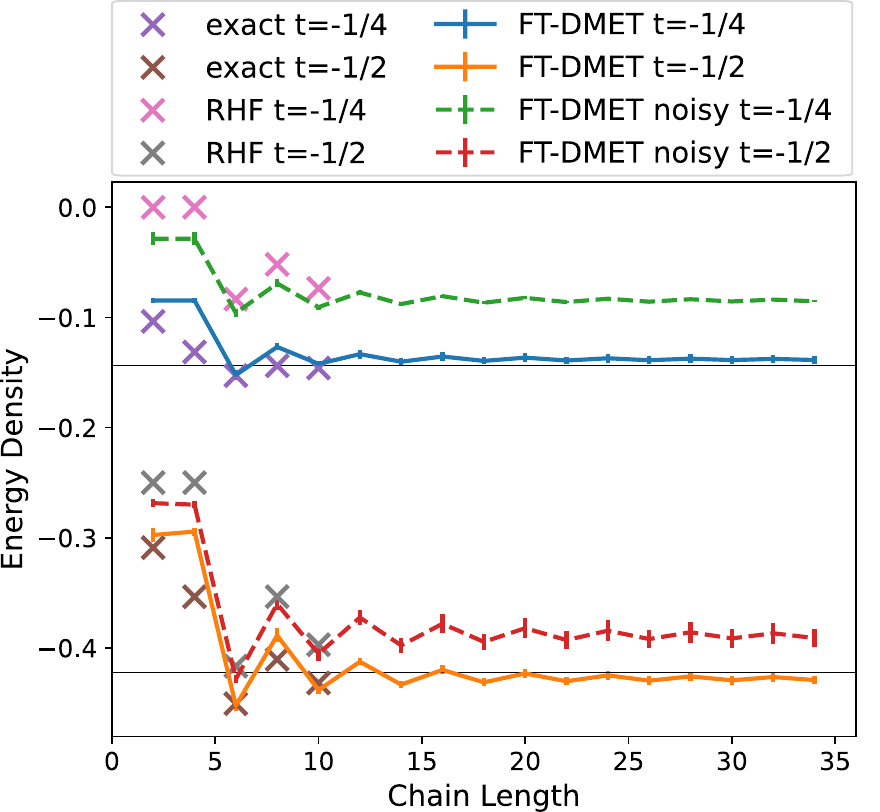}
\caption{\label{fig:FermiHubbardDMET} The result for the one-dimensional Fermi-Hubbard model at half filling using FT-DMET with the noisy data (dashed) and clean data (solid) compared to restricted Hartree-Fock (RHF). The black horizontal lines indicate the exact solutions in the thermodynamic limit,  the error bars where obtained by training five neural networks on the same training data.}
\end{figure}

\begin{figure}
\includegraphics[width=.95\linewidth]{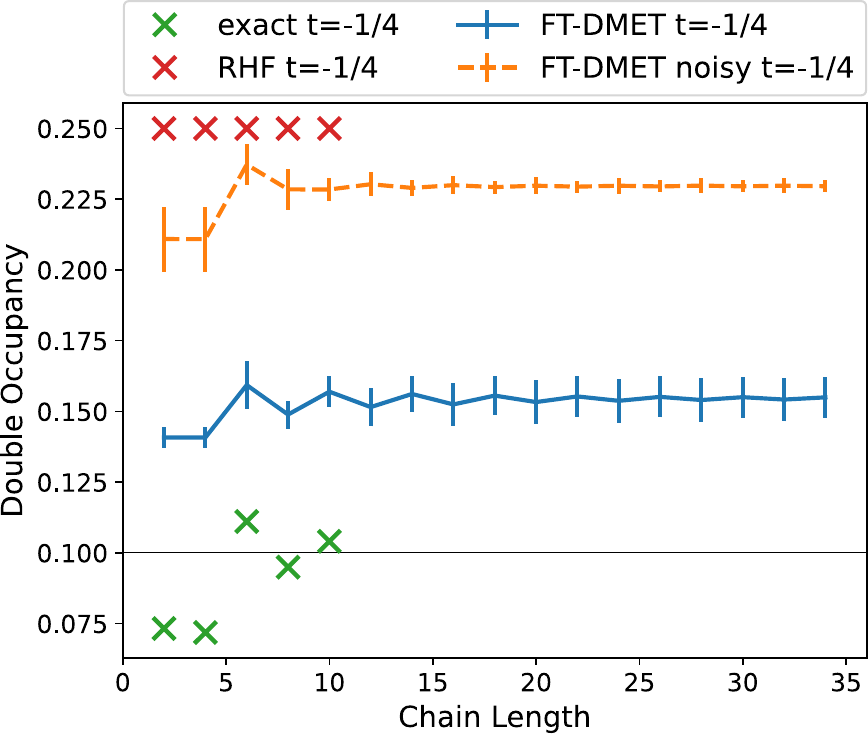}
\caption{\label{fig:FermiHubbardDMET_occ} The double occupancy $\langle n_{i\uparrow} n_{i\downarrow} \rangle$ at $t=-1/4$, where again the clean (solid) gives a more substantial improvement over the noisy (dashed) data. The neural net was trained five times on the same data to gauge the error-bars.}
\end{figure}

To generalize this functional to different specific model configurations, it is assumed to be only one-dimensional with nearest-neighbouring hopping. These restrictions improve comparability to exact results, and can easily be extended to the case of multiple dimensions, of oddly shaped lattice geometries, and of non-uniform hopping strengths or potentials. The resulting energy is in good agreement with the exact result at modest chain-lengths, and with the analytical result for the thermodynamic limit for half-filling (Fig.~\ref{fig:FermiHubbardDMET}). The exact result for the infinite chain limit was obtained using the Bethe-Ansatz \cite{Lieb1968, Shiba1972}. Alongside the energy, the probability of double occupancy $\langle n_{i\uparrow} n_{i\downarrow} \rangle$ is plotted in Fig.~\ref{fig:FermiHubbardDMET_occ}. The functional formalism gives direct access to this information, as we can easily subtract the whole kinetic energy contribution from the energy, and are left with this value times the interaction strength. This again shows significant improvements over Hartree-Fock, whose solution implicitly assumes a maximal occupation. For this metric, even the noisy data also shows significant improvements over Hartree-Fock. A small comment regarding chain lengths of 4, 8, 12 and so on is required. Since the eigenvalues of the hopping-term matrix are not clearly distinguishable, a slight offset was introduced to increase the negativity on every other hopping-term by $0.01$, thus removing this indistinguishability. This minor offset does not affect the energy outcome notably.

\begin{figure}[ht]
\includegraphics[width=.95\linewidth]{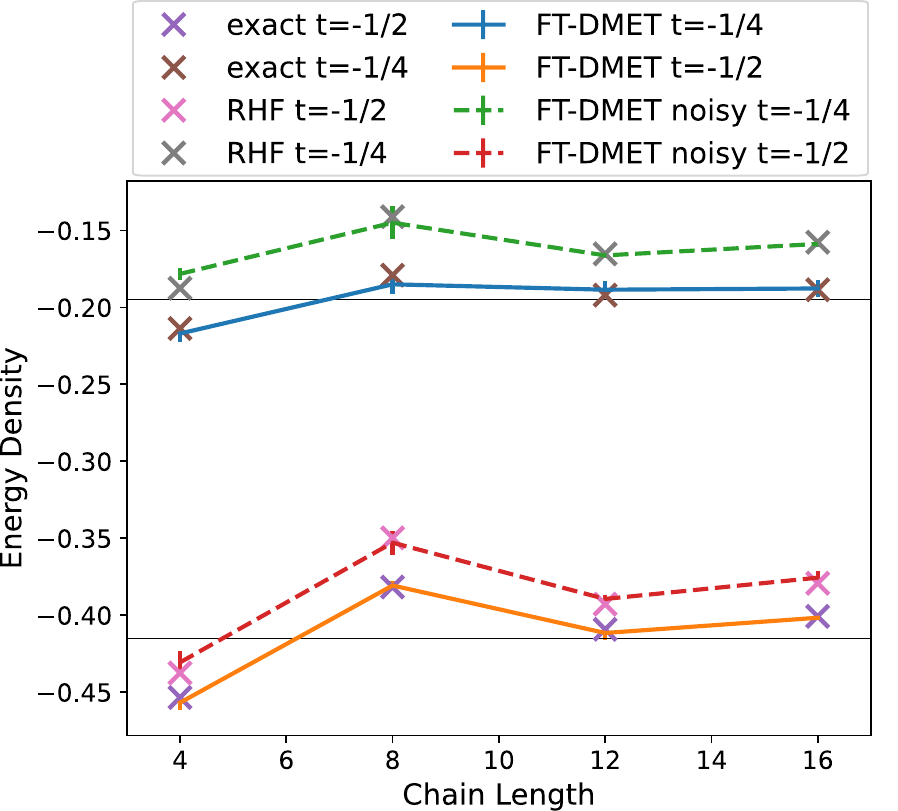}
\caption{\label{fig:FermiHubbardQuarterDMET} Using the FT-DMET scheme, the result for the one-dimensional Fermi-Hubbard model at one-quarter filling with both clean (solid) and noisy data (dashed). The black line indicates the exact result for the infinite system size.}
\end{figure}

One advantage of this method is the ability to reuse the same DNN-functional for different cases as shown in Fig.~\ref{fig:FermiHubbardQuarterDMET}. In there we show the results for one quarter and three quarter fillings. Particularly, for the one quarter filling case the noisy data is worse than the Hartree-Fock result, while the result for the three quarter filling is on average an improvement. This behavior is caused by the interaction energy being a more important factor for the three quarter filling, as it consists of more particles. This limit, in particular, is one for which the Hartree-Fock approach struggles as it overvalues the interaction energy contribution substantially, while the FT-DMET scheme is able to deal with it more accurately (see lower panel of Fig.~\ref{fig:FermiHubbardDMET}).

\begin{figure}[ht]
\includegraphics[width=.95\linewidth]{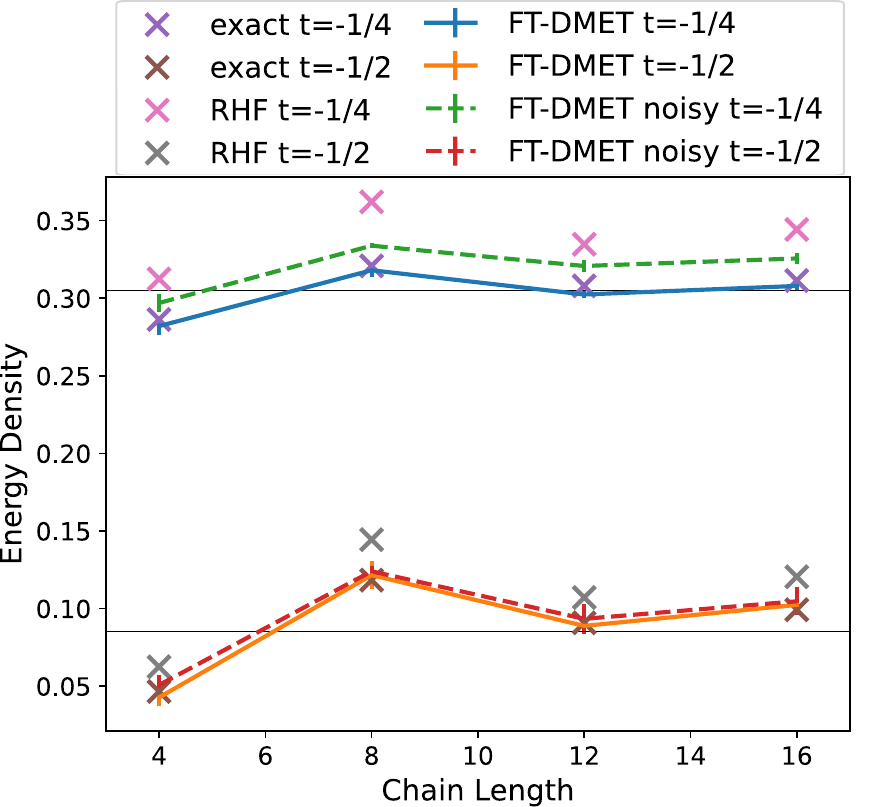}
\caption{\label{fig:FermiHubbardThreeQuarterDMET} With the FT-DMET scheme, the result for the one-dimensional Fermi-Hubbard model at three-quarter filling with both clean (solid) and noisy data (dashed) as well as the exact infinite chain limit in black.}
\end{figure}

\newpage

\subsubsection{\label{subsec:TwoBandFermiHubbard} Fermions: Two-Band Fermi-Hubbard Model}

\begin{figure*}[ht]
\centering
\includegraphics[width=.95\linewidth]{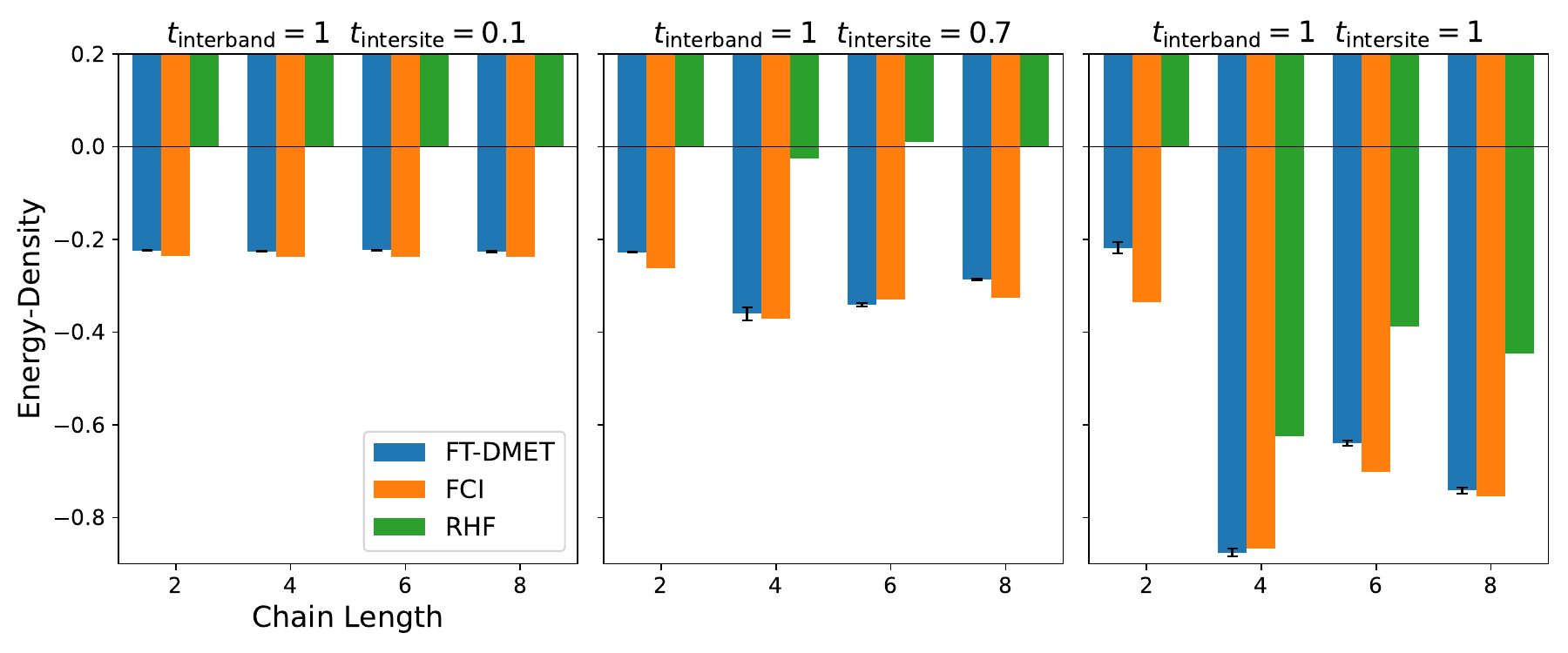}
\caption{\label{fig:TwoBandFermiHubbardQCData_reducedVariationality} The energy-density for the two-band Fermi-Hubbard model as defined in Eq.~\eqref{equ:HamiltonianTwoBandFermiHubbard} for different chain lengths and intersite hopping terms. The error was obtained by training the DNN-functional on the same training data five times with different seeds.}
\end{figure*}

As an example of a larger model, we use the two-band Fermi-Hubbard model with a fragment and bath size of 2 orbitals each. The general Hamiltonian analyzed here is of the form
\begin{align}\label{equ:HamiltonianTwoBandFermiHubbard}
    H = &\sum_i \bigg( w_0 \sum_{j\in \{0,1\}} n_{i j \uparrow} n_{i j \downarrow}  \\ 
    &+ w_1 \sum_{j\in \{0,1\}} \sum_{\sigma} n_{i j {\sigma}} n_{i \bar{j} \bar{\sigma}} \bigg) \nonumber \\
    &+ t_{\text{interband}} \sum_i \sum_{j\in \{0,1\}} \sum_{\sigma} \hat{a}^\dagger_{ij\sigma} \hat{a}_{i\bar{j}\bar{\sigma}} \nonumber \\
    &+ t_{\text{intersite}} \sum_{\langle i, k\rangle} \sum_{j\in \{0,1\}} \sum_{\sigma} \hat{a}^\dagger_{ij\sigma} \hat{a}_{kj\bar{\sigma}} \nonumber ,
\end{align}
where for simplicity we assume the interaction on both sites of the two-band model to be identical, with $w_0=3$, and the interband interaction $w_1=1$. Additionally, the intersite hopping and the interband hopping strengths are two distinct and independently tunable values. The fragments are taken here as the interacting two-orbital chunks. To reduce the number of required input parameters of the DNN, thus reducing the number of required training data, we fixed $t_{\text{interband}}$ to 1. In addition, we focused on intersite hoppings between 0 and 1, as for those the fragments are the dominant units in the mean-field solution leading to a single-body Hamiltonian where the hopping terms between bath and fragment have a clear and fixed dependence on one-another. This can be exploited in the DNN-functional by defining a new variable which is the combination of these hopping terms. While this reduces the universality of the DNN-functional, such simplification can be used in many systems. As an example, a $(\text{H}_2)_2$ combination of two $\text{H}_2$ molecules does have the same mean-field 1-RDM for any distance leading to the same projection onto the embedding Hamiltonian. 

In Fig.~\ref{fig:TwoBandFermiHubbardQCData_reducedVariationality}, we show a markedly improved accuracy over the RHF result, which is the starting point for the FT-DMET scheme. The result for small intersite hoppings (here $t_{\text{intersite}}=0.1$) is particularly accurate as, in such case, the fragments serve as a particularly good approximation for the whole system. For larger intersite hoppings, some of the information encoded in the entanglement between the fragments is inevitably lost. However, the results are still promising considering the low computational effort compared to the FCI calculation for the longer chains. The proposed FT-DMET scheme would be entirely capable of describing larger systems without drastic increase the computational complexity. The computational time between the different FT-DMET results is only changing marginally. Nevertheless, the minimization process of Eq.~\eqref{equ:FTDMETMinimizationWithDNN} is faster for the more localized examples, as the particle number distribution in the bath is heavily focused on one of the two sites. We would like to emphasize that the here learned functional is only applicable to one specific type of interaction term. This could be extended using mean field correction of the interaction energy.

\section{\label{sec:ConclusionOutlook}Conclusion and Outlook}

\begin{figure*}
\footnotesize
    \centering
    \scalebox{1.1}{
   \begin{tikzpicture}[node distance=0.2cm, every node/.style={rounded corners, align=center}, arrow/.style={-{Stealth}}]
   \scriptsize
        \node (FirstNode) {Fix Interaction $W$ of the Hamiltonian Eq.~\eqref{equ:GenericHamiltonian}};
        \node (SecondNode) [below=of FirstNode, yshift=-0.4cm] {Randomize Many \\ Single-body-Hamiltonians $h$ Eq.~\eqref{equ:GenericHamiltonian}};
        \draw[arrow] (FirstNode) -- (SecondNode);
        \node (ThirdNode) [below=of SecondNode, yshift=-0.4cm] {Solve the Full Hamiltonians (Alg.~\ref{alg:VQE}) \\ and Calculate Functional Values Eq.~\eqref{equ:RDMFunctionalCalculation}};
        \draw[arrow] (SecondNode) -- (ThirdNode);
        \node (FourthNode) [below=of ThirdNode, yshift=-0.4cm] {Learn Functional $\mathcal{F}_{\text{DNN}}[\gamma]$ \\ with DNN};
        \draw[arrow] (ThirdNode) -- (FourthNode);
        \node (FifthNode) [below=of FourthNode, yshift=-.8cm] {Get the 1-RDM $\gamma$ \\ of the Entire System \\ with a Mean-Field Method};
        \draw[arrow, line width=1mm, color=orange]  (2.8,-4.6) -- (2.8,-5.25);
        \node (SixthNode) [below=of FifthNode, yshift=-0.4cm] {Create Embedded Problem \\ with the Projector Eq.~\eqref{equ:BosonicProjector} / Eq.~\eqref{equ:fermionicProjector} \\ Applied to the Hamiltonian Eq.~\eqref{equ:CreateEmbeddingHamiltonian}};
        \draw[arrow] (FifthNode) -- (SixthNode);
        \node (SeventhNode) [below=of SixthNode, yshift=-0.4cm] {Solve Embedded Problem \\ with the Functional Eq.~\eqref{equ:FTDMETMinimizationWithDNN} };
        \draw[arrow] (SixthNode) -- (SeventhNode);
        \node (EigthNode) [below=of SeventhNode, yshift=-0.4cm] {Calculate the Real Energy by \\ Summing up the Fragments Eq.~\eqref{equ:FTDMETEnergy} };
        \draw[arrow] (SeventhNode) -- (EigthNode);
        \draw[draw=teal, line width=2pt, rounded corners] (-3.25, .5) rectangle (8.5,-4.55);
        \node at (6.75,-1) {\makecell{\color{teal} \textbf{Execute} \\ \color{teal} \textbf{Once}\\ \color{teal} \textbf{Including} \\ \color{teal} \textbf{ Use of QPUs}} \\ \color{teal} \textbf{(parallelizable)} };
        \draw[draw=olive, line width=2pt, rounded corners] (-3.25, -5.25) rectangle (8.5,-11.2);
        \node at (6.75,-9.25) {{\makecell{\color{olive} \textbf{Reuse} \\ \color{olive} \textbf{Multiple} \\ \color{olive} \textbf{Times} \\ \color{olive} \textbf{Only Using} \\ \color{olive} \textbf{ CPUs}}}};
        
        \node at (3.75,-2.65) {\scalebox{.5}{\begin{tikzpicture}[
  wire/.style={black, line width=0.8pt},
  gate/.style={draw=black, fill=gray!20, rounded corners=4pt, minimum width=.5cm, minimum height=1cm},
  qubit/.style={anchor=east},
  meas/.style={draw=black, fill=gray!10, minimum size=6pt, inner sep=2pt}
]

% Define vertical spacing between qubits
\def\yspacing{.5};

% Qubit lines
\foreach \i in {0,1} {
  \draw[wire] (-.25,\i*\yspacing) -- (2,\i*\yspacing);
  \node[qubit] at (-0.2,\i*\yspacing) {$|0\rangle$};
};

% Multi-qubit gate box
\node[gate] (u) at (.75,\yspacing-.25) {$U(\boldsymbol\theta)$};

% Measurement symbols
\foreach \i in {0,1} {
  \node[meas] at (2.,\i*\yspacing) {$\text{M}$};
};

\end{tikzpicture}}};

        \node at (3.75,-5.95) {
        \scalebox{.3}{
        \begin{tikzpicture}
            \foreach \Point in {(0,0), (1,0), (2,0), (3,0), (0,-1), (3,-1),(0,-2), (1,-2), (2,-2), (3,-2)}{
            \node at \Point {\Large \textbullet};
        }
            \foreach \Point in {(1,-1), (2,-1)}{
            \node at \Point {\color{black} \Large \textbullet};
        }
        
            \draw[thick] (-0.5,0.02) -- (-0.2,0.02);
            \draw[thick] (0.2,0.02) -- (0.8,0.02);
            \draw[thick] (1.2,0.02) -- (1.8,0.02);
            \draw[thick] (2.2,0.02) -- (2.8,0.02);
            \draw[thick] (3.2,0.02) -- (3.5,0.02);
        
            \draw[thick] (-0.5,-0.98) -- (-0.2,-0.98);
            \draw[thick] (0.2,-0.98) -- (0.8,-0.98);
            \draw[thick] (1.2,-0.98) -- (1.8,-0.98);
            \draw[thick] (2.2,-0.98) -- (2.8,-0.98);
            \draw[thick] (3.2,-0.98) -- (3.5,-0.98);
        
            \draw[thick] (-0.5,-1.98) -- (-0.2,-1.98);
            \draw[thick] (0.2,-1.98) -- (0.8,-1.98);
            \draw[thick] (1.2,-1.98) -- (1.8,-1.98);
            \draw[thick] (2.2,-1.98) -- (2.8,-1.98);
            \draw[thick] (3.2,-1.98) -- (3.5,-1.98);
        
            \draw[thick] (0,0.52) -- (0,0.22);
            \draw[thick] (0,-0.78) -- (0,-0.18);
            \draw[thick] (0,-1.78) -- (0,-1.18);
            \draw[thick] (0,-2.48) -- (0,-2.18);
        
            \draw[thick] (1,0.52) -- (1,0.22);
            \draw[thick] (1,-0.78) -- (1,-0.18);
            \draw[thick] (1,-1.78) -- (1,-1.18);
            \draw[thick] (1,-2.48) -- (1,-2.18);
        
            \draw[thick] (2,0.52) -- (2,0.22);
            \draw[thick] (2,-0.78) -- (2,-0.18);
            \draw[thick] (2,-1.78) -- (2,-1.18);
            \draw[thick] (2,-2.48) -- (2,-2.18);
        
            \draw[thick] (3,0.52) -- (3,0.22);
            \draw[thick] (3,-0.78) -- (3,-0.18);
            \draw[thick] (3,-1.78) -- (3,-1.18);
            \draw[thick] (3,-2.48) -- (3,-2.18);
        \end{tikzpicture}}};

        \node at (3.75,-4.05) {\scalebox{.15}{
\begin{tikzpicture}[
  node/.style={circle, draw=black, fill=gray!20, minimum size=8mm, inner sep=0},
  layer label/.style={gray!60, font=\small},
  connection/.style={black},
  >=Stealth
]

% Layer positions
\def\layersep{2.5cm}
\def\nodesep{1.2cm}

% Input layer
\foreach \i in {0,1,2}
  \node[node] (I\i) at (0,-\i*\nodesep) {};

% Hidden layer
\foreach \i in {0,1,2,3}
  \node[node] (H\i) at (\layersep,-\i*\nodesep + 0.6cm) {};

% Output layer
\foreach \i in {1}
  \node[node] (O\i) at (2*\layersep,-\i*\nodesep - 0.cm) {};

% Connections: Input -> Hidden
\foreach \i in {0,1,2}
  \foreach \j in {0,1,2,3}
    \draw[connection] (I\i) -- (H\j);

% Connections: Hidden -> Output
\foreach \i in {0,1,2,3}
  \foreach \j in {1}
    \draw[connection] (H\i) -- (O\j);

\end{tikzpicture}}};
        
        \node at (3.75,-7.65) {
        \scalebox{.3}{
        \begin{tikzpicture}
            \foreach \Point in {(0,0), (1,0), (2,0), (3,0), (0,-1), (3,-1),(0,-2), (1,-2), (2,-2), (3,-2)}{
            \node at \Point {\Large \textbullet};
        }
            \foreach \Point in {(1,-1), (2,-1)}{
            \node at \Point {\color{red} \Large \textbullet};
        }
        
            \draw[thick] (-0.5,0.02) -- (-0.2,0.02);
            \draw[thick] (0.2,0.02) -- (0.8,0.02);
            \draw[thick] (1.2,0.02) -- (1.8,0.02);
            \draw[thick] (2.2,0.02) -- (2.8,0.02);
            \draw[thick] (3.2,0.02) -- (3.5,0.02);
        
            \draw[thick] (-0.5,-0.98) -- (-0.2,-0.98);
            \draw[thick] (0.2,-0.98) -- (0.8,-0.98);
            \draw[thick] (1.2,-0.98) -- (1.8,-0.98);
            \draw[thick] (2.2,-0.98) -- (2.8,-0.98);
            \draw[thick] (3.2,-0.98) -- (3.5,-0.98);
        
            \draw[thick] (-0.5,-1.98) -- (-0.2,-1.98);
            \draw[thick] (0.2,-1.98) -- (0.8,-1.98);
            \draw[thick] (1.2,-1.98) -- (1.8,-1.98);
            \draw[thick] (2.2,-1.98) -- (2.8,-1.98);
            \draw[thick] (3.2,-1.98) -- (3.5,-1.98);
        
            \draw[thick] (0,0.52) -- (0,0.22);
            \draw[thick] (0,-0.78) -- (0,-0.18);
            \draw[thick] (0,-1.78) -- (0,-1.18);
            \draw[thick] (0,-2.48) -- (0,-2.18);
        
            \draw[thick] (1,0.52) -- (1,0.22);
            \draw[thick] (1,-0.78) -- (1,-0.18);
            \draw[thick] (1,-1.78) -- (1,-1.18);
            \draw[thick] (1,-2.48) -- (1,-2.18);
        
            \draw[thick] (2,0.52) -- (2,0.22);
            \draw[thick] (2,-0.78) -- (2,-0.18);
            \draw[thick] (2,-1.78) -- (2,-1.18);
            \draw[thick] (2,-2.48) -- (2,-2.18);
        
            \draw[thick] (3,0.52) -- (3,0.22);
            \draw[thick] (3,-0.78) -- (3,-0.18);
            \draw[thick] (3,-1.78) -- (3,-1.18);
            \draw[thick] (3,-2.48) -- (3,-2.18);

            \draw[thick,dashed,color=red] (0.5,-0.48) -- (2.5,-0.48);
            \draw[thick,dashed,color=red] (0.5,-0.48) -- (0.5,-1.48);
            \draw[thick,dashed,color=red] (0.5,-1.48) -- (2.5,-1.48);
            \draw[thick,dashed,color=red] (2.5,-1.48) -- (2.5,-0.48);
        \end{tikzpicture}}};

        \node at (3.75,-9.25) {
        \scalebox{.3}{
        \begin{tikzpicture}
            \foreach \Point in {(7.5,-0.48), (8.5,-0.48)}{
                \node at \Point {\Large \color{red} \textbullet};
            }
            \foreach \Point in {(7.5,-1.48), (8.5,-1.48)}{
                \node at \Point {\Large \color{blue} \textbullet};
            }
        
            \draw[thick] (7.7,-0.48) -- (8.3,-0.48);
            \draw[thick] (8.5,-0.68) -- (8.5,-1.28);
            \draw[thick] (7.7,-1.48) -- (8.3,-1.48);
    \draw[thick] (7.5,-0.68) -- (7.5,-1.28);
        \end{tikzpicture}}};

        \node at (3.75,-10.67) {
        \scalebox{2.}{$E_0$}};

         \draw [draw, rounded corners, align=center, fill=gray, opacity=0.25, font=\footnotesize, line width=0pt](-3.1, 0.35) rectangle (5.,-0.25);
        
         \draw [draw, rounded corners, align=center, fill=gray, opacity=0.25, font=\footnotesize, line width=0pt](-3.1, -.86) rectangle (5.,-1.65);
        
         \draw [draw, rounded corners, align=center, fill=teal, opacity=0.25, font=\footnotesize, line width=0pt](-3.1, -2.26) rectangle (5.,-3.05);        
        
         \draw [draw, rounded corners, align=center, fill=gray, opacity=0.25, font=\footnotesize, line width=0pt](-3.1, -3.71) rectangle (5.,-4.4);

         \draw [draw, rounded corners, align=center, fill=gray, opacity=0.25, font=\footnotesize, line width=0pt](-3.1, -5.4) rectangle (5.,-6.5);
         \draw [draw, rounded corners, align=center, fill=gray, opacity=0.25, font=\footnotesize, line width=0pt](-3.1, -7.13) rectangle (5.,-8.2);
         \draw [draw, rounded corners, align=center, fill=gray, opacity=0.25, font=\footnotesize, line width=0pt](-3.1, -8.9) rectangle (5.,-9.6);
         \draw [draw, rounded corners, align=center, fill=gray, opacity=0.25, font=\footnotesize, line width=0pt](-3.1, -10.28) rectangle (5.,-11.05);
         
         \node at (6.75,-7.25) {
         \scalebox{.4}{
            \begin{tikzpicture}[scale=1, every node/.style={inner sep=0, outer sep=0}]
            
            % Laptop base
            \fill[gray!70, sharp corners] (-3,-0.2) rectangle (3,0);
            \fill[gray!50, sharp corners] (-3,0) -- (3,0) -- (2.5,0.3) -- (-2.5,0.3) -- cycle;
            
            % Screen
            \fill[gray!40, sharp corners] (-2.5,0.3) rectangle (2.5,3);
            \fill[black!80!gray, sharp corners] (-2.4,0.4) rectangle (2.4,2.9); % screen content
            
            % Camera
            \fill[black!60, sharp corners] (0,2.85) circle (0.05);
            \end{tikzpicture}}};
            
        \node at (6.75,-3) {\scalebox{.4}{
\begin{tikzpicture}
% Top mounting structure
\filldraw[fill=gray!30, draw=gray!70, line width=2pt, sharp corners] 
    (-3, 4.5) rectangle (3, 4.2);

% Main support structure
\filldraw[fill=gold!20, draw=gold!100!black, line width=1.5pt, sharp corners] 
    (-2.5, 4.2) -- (-2.5, 3.8) -- (2.5, 3.8) -- (2.5, 4.2);

% The iconic golden hanging cylinders in a chandelier pattern
\foreach \x/\y/\h/\r in {
    0/3.8/2.5/0.3,           % Center - longest
    -1.2/3.8/2.2/0.25,       % Inner left
    1.2/3.8/2.2/0.25,        % Inner right  
    -2/3.8/1.8/0.2,          % Outer left
    2/3.8/1.8/0.2,           % Outer right
    -0.6/3.8/2.0/0.22,       % Mid left
    0.6/3.8/2.0/0.22         % Mid right
} {
    % Golden cylinder body
    \filldraw[fill=gold!10, draw=gold!100!black, line width=1pt, sharp corners] 
        (\x-\r, \y) rectangle (\x+\r, \y-\h);
    
    % Connecting cable from top
    \draw[thick, gold!60, sharp corners] (\x, 4.2) -- (\x, \y);
    
    % Golden bands/ridges on cylinders
    \foreach \i in {0.3, 0.6, 0.9} {
        \draw[gold!90!black, line width=0.8pt, sharp corners] 
            (\x-\r, \y-\i) -- (\x+\r, \y-\i);
    }
}

% The quantum chip at the very bottom (center cylinder)
\filldraw[fill=gold!50, draw=gold!80!black, line width=2pt, sharp corners]     (-0.2, 1.3) rectangle (0.2, 1.0);
\end{tikzpicture}}};
        
    \end{tikzpicture}}
    \caption{The workflow of FT-DMET is illustrated here using VQE to generate the training data.}
    \label{fig:WorkflowDetailed}
\end{figure*}
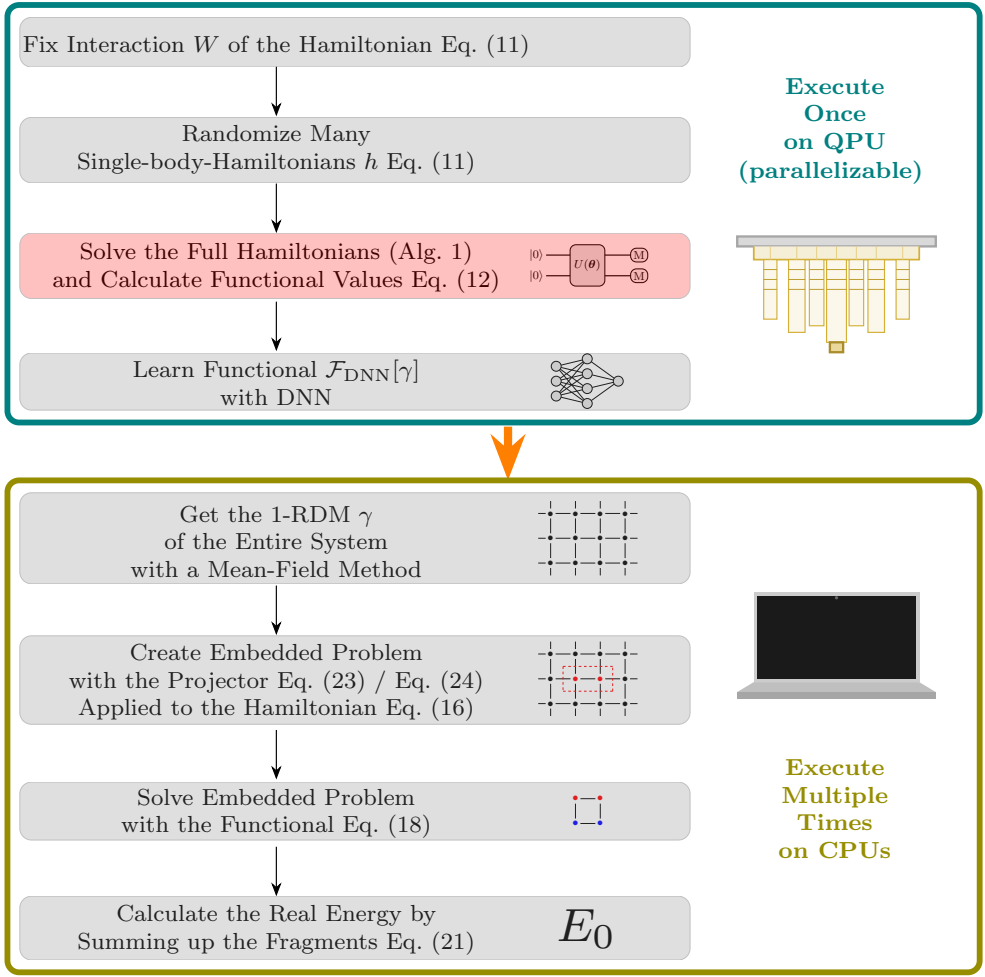

Here we propose and apply a quantum-ready workflow that combines RDM-FT with DMET to obtain the universal functional, and to potentially provide deeper understanding of systems with significant electronic correlations. In particular, we propose to use machine learning in order to obtain a DNN-functional inspired by the RDM-FT formalism, a route that can be applied across multiple chemical and condensed matter systems. This creates the potential for a cumulative advantage in quantum computing applications, depending on the efficiency of the quantum algorithm used in the proposed workflow. This method as it was implemented for this work is depicted in detail within Fig.~\ref{fig:WorkflowDetailed}. After initially determining the class of systems under investigation by fixing the interaction, VQE solves many full Hamiltonians with the single-body Hamiltonian-terms being randomized. This is then fed into the here defined DNN-Functional. In order to then find the ground state energy of a specific system, it is first solved on a mean-field level to then build a DMET-embedded system which is solved with the DNN-functional.

A significant practical benefit of our method is its reduced dependence on QPUs for general quantum chemical calculations. Given that quantum hardware is likely to remain a limited resource in the near future, this efficiency is particularly valuable. Furthermore, our approach offers scalability, as the training data it requires can be generated via parallel and independent processes.

% albeit large amounts of said training data is needed to accurately describe larger fragment-bath systems. (Comment Daniel: clearer mentions of this issue are already included later on in the conclusions.)

By incorporating the DMET framework we have enhanced the functional's universality, which justifies the large initial QPU-cost incurred to obtaining the training data. The proposed approach delivers accurate results while maintaining computational efficiency, comparing favorably with existing quantum chemical methods of similar computational cost. In particular, for lattice models the only scaling components are the mean-field method and the diagonalization of the environment's 1-RDM, both expected to scale cubic with respect to system size. When extending this scheme to molecular systems, the cost is expected to mostly depend on the implementation of the orbital localization.

\normalsize

This work can be expanded to orbital-based quantum chemical calculations. For the lattice models discussed here, the scaling is $\mathcal{O}(N^3)$. However, depending on the specific systems, larger and thus more computationally intensive embedded systems need to be solved for an accurate result. An additional requirement for extending this scheme is the orthogonalisation of the orbitals \cite{Lwdin1950, Sun2014, Aquilante2006}. This requires care in the context of orbital selection as the post-orthogonalisation two-electron repulsion integrals need to be identical (up to a rescaling factor) to allow for usage of the same DNN-Functional. 

As larger quantum computers become a reality, our workflow could be adapted to include more advanced quantum algorithms. These would better capture complex entanglement effects of larger systems. Such advances would particularly benefit the FT-DMET calculations, enabling more precise measurements of interaction energies. Nevertheless, even for relatively small quantum computers likely to be available in the near future, the flexibility of our workflow allows the inclusion of techniques that allow scaling up current calculations. Particularly issues that one needs to address if VQE continues to be used are the "barren plateau" problem \cite{Wang2021, Uvarov2021}, and the large number of training data required. In such case one could modify the proposed workflow to combine the basic FT-DMET framework with quantum subspace expansion techniques, as demonstrated by Yoshioka et al. \cite{Yoshioka2022}. Additionally, for larger system more specialized neural networks, like convolutional layers, could prove a powerful extension of this work, as these put more emphasis on local dependencies e.g.~between the densities \cite{Koridon24}. This combination could also enable the study of low-lying excited states, an area where ensemble RDM-FT already provides a solid theoretical foundation \cite{Liebert2021}.

\begin{acknowledgments}
We wish to acknowledge the support by the Quantum Computing Initiative of the DLR funded by the Bundesministerium fuer Forschung, Technologie, und Raumfahrt (BMFTR) (https://qci.dlr.de/quanticom, https://qci.dlr.de/qcmineral). M.J.U. is funded by the DLR Quantum Computing Initiative through the Quantum Fellowship Programme. No specific grant number is available. The authors gratefully acknowledge the computational and data resources provided through the joint high-performance data analytics (HPDA) project “terrabyte” of the German Aerospace Center (DLR) and the Leibniz Supercomputing Center (LRZ).

\textbf{Data Availability}\newline
The data and code used within this work is available upon reasonable request.
\end{acknowledgments}

\textbf{Author Contributions} \newline
M.J.U. implemented the code, ran the simulations and wrote the manuscript in collaboration with M.L. and D.B.-Y. who were involved in planning, discussing and analyzing the result. M.L., D.B.-Y. and M.S. reviewed the manuscript. LLMs were used in the creation of the code.

\textbf{Competing Interests} \newline
The authors declare no competing interests.

\appendix

\section{\label{appendix:SimilarDensitiesSimilarWavefunctions} Similar Densities Imply Similar Wave-functions}

In this section, it is argued more rigorously that under certain conditions, most notably the ground state being non-degenerate, then similar functional variables (e.g.~the density) imply similar ground state wave functions. Put more formally, the 1-to-1 relation proposed by the Hohenberg-Kohn theorem \cite{Hohenberg1964} between the ground state and the density (or 1-RDM or other functional variable) is continuous. The argument requires working with finite dimensional functional variables (like the setting in the work above). For the infinite dimensional case with the true density functional, the following argument will not hold \cite{Kvaal2014}.

The argument is split up into two parts. \textit{First}, it is shown that the relation mapping the functional variable onto the external potential is continuous. \textit{Second}, the ground state varies only infinitesimally if the Hamiltonian is varied infinitesimally due to the adiabatic theorem. In this argument, some concepts of convex analysis are required \cite{Rockafellar1996} since functional and energy are connected via a Legendre-Fenchel transformation \cite{Schilling2018}.

The following argument is generally applicable to any functional theory. To that end, the Hamiltonian is of the general form
\begin{equation}
  \hat{H} = \hat{H}_0 + v_{\text{ext}} \hat{H}_1
\end{equation}
with $v_{\text{ext}}$ being some real vector. $\hat{H}_0$ is described by the functional while the energy contribution of the rest can be found via a scalar product $\langle v_{\text{ext}}, \rho \rangle$ with $\rho = \langle \hat{H}_1 \rangle$ being referred to as the generalized density or functional variable (e.g.~the reduced density matrix).

First, one is able to show that the functional variable $\rho$ and $v_{\text{ext}}$ are connected continuously. As it turns out, it is required for the ground state energy to be strictly concave. This is in general a reasonable assumptions if the Hohenberg-Kohn theorem is valid as if it were non-strictly concave, it would have a line segment meaning that the same generalized density is associated with multiple external potentials $\rho = \partial_{v_{\text{ext}}} E_0(v_{\text{ext}})$ thereby destroying the one-to-one relation which is at the heart of Hohenberg-Kohn. This is of course in general entirely possible, in quantum chemistry it can however be assumed not being so and conditions can be found to more general cases \cite{Xu2022}. Note that we restrict the functional to the domain on which it is well-defined i.e.~finite.

\begin{lemma}
  If the ground state energy is strictly concave, then a function $f:\rho\mapsto v_{\text{ext}}$ relating the generalized density to its associated external potential connected via a Legendre transform is continuous.
\end{lemma}
\begin{proof}
  For the following proof, a few general statements regarding the functional are of help. Due to Rellich's theorem \cite{Rellich1937}, the ground state energy of finite systems needs to be continuous. According to Rockafellar \cite{Rockafellar1996}, the functional as defined here - which is a Legendre-(Fenchel-)transform of the ground state energy - is necessarily convex and lower semi-continuous. Since the domain of the functional is defined as only the points on which it is well-defined and finite, this statement can be extended to continuousness of the functional. The reason for this is that any non-continuos point would be associated with an infinite slope meaning the functional cannot be well-defined on all sides of such a point while keeping convexity.

  Now due to the relation via a Legendre-Fenchel-transform, the external potential and the functional are related via
  \begin{equation*}
    v_{\text{ext}}(\rho) = \partial_{\rho} \mathcal{F}[\rho]
  \end{equation*}
  where $\partial_{\rho} \mathcal{F}[\rho]$ is the subdifferential which is also well-defined even if $\mathcal{F}$ is non-differentiable \cite{Rockafellar1996}. If the functional is continuously differentiable, then the proof would be finished as then $v_{\text{ext}}(\rho)$ would trivially be continuous. Supposing the functional being non-differentiable inevitably implies a kink i.e.~a situation where $\partial_{\rho} \mathcal{F}[\rho]$ is a non-singular set at some $\rho_0$. This would however imply the ground state energy being linear there or in other words the density $\rho_0$ being associated with all $v_{\text{ext}} \in \partial_\rho \mathcal{F} |_{\rho \rightarrow \rho_0}$ thus contradicting the Hohenberg-Kohn theorem.
\end{proof}

Secondly, using the adiabatic theorem, we can directly see that a change in the external potential changes the ground state wave function in a continuous manner. In particular, if the energy gap between the ground and lowest excited energy is nonzero, then the relation $\ket{\Psi_0(v_{\text{ext}})}$ is continuous or, in other words, an infinitesimal change in the external potential leads to an infinitesimal change in the ground state wave function. To make this more explicit, think of the Hamiltonian as
\begin{equation}
  \hat{H}\left(v_{\text{ext}}(t)\right) = \hat{H}_0 + v_{\text{ext}}(t) \hat{H}_1
\end{equation}
and that the external potential depends continuously but arbitrarily on the time. Then the adiabatic approximation is
\begin{equation}
  \ket{\Psi_{\text{adiabatic}}(t)} = e^{i \theta_n(t)} e^{i \gamma_n(t)} \ket{\Psi_n(t)}
\end{equation}
which clearly implies the above statement \cite{Sakurai2020}. Taken in conjunction with the earlier lemma, we find that if the energy is strictly concave and non-degenerate, then the relation
\begin{equation}
  \ket{\Psi_0} = \ket{\Psi_0(v_{\text{ext}})} = \ket{\Psi_0(v_{\text{ext}}(\rho))}
\end{equation}
is continuous. For the work above, this implies that if the ground state is non-degenerate and the Hohenberg-Kohn theorem is valid, the wave-function does not change significantly when changing the density i.e.~small errors in the quantum computer still give results close to the exact one.

\section{\label{appendix:rank1matrix} The Eigenvalues of the Bosonic Mean-Field 1-RDM}

The 1-RDM constructed for the bosons on a mean-field level $\gamma_{MF}$ has a very peculiar eigenvalue structure with all but one eigenvalue being zero. This can easily be shown. Any matrix constructed by the outer product of the same vector used $n$-times gives a matrix of the form
\begin{align}
 \gamma_{MF} &= (v_0, v_0, v_0, ...) \cdot (v_0, v_0, v_0, ...)^T \\
 &= \begin{pmatrix} n \left(v_0^{(0)}\right)^2 & n  v_0^{(0)} v_0^{(1)} & n  v_0^{(0)} v_0^{(2)} & \\
 n  v_0^{(1)} v_0^{(0)} & n  \left(v_0^{(1)}\right)^2 & n  v_0^{(1)} v_0^{(2)} & ... \\
 n  v_0^{(2)} v_0^{(0)} & n  v_0^{(2)} v_0^{(1)} & n  \left(v_0^{(2)}\right)^2 &  \\
  & ... &  &  \end{pmatrix} \nonumber
\end{align}
which can be rewritten as the outer product of the vector $v=(\sqrt{n} v_0^{(0)}, \sqrt{n} v_0^{(1)}, \sqrt{n} v_0^{(2)}, ... )^T$ with itself implying that the matrix $\gamma_{MF}$ is of rank one. Removing the columns and rows corresponding to the fragment does not change that fact. Without loss of generality, it is assumed that the $m$ last columns and rows are the ones that are removed. It is clear to see that this does not affect the other elements of $\gamma_{MF}$. Thus, the rank of $\gamma_{MF}^{\text{env}}$ is still one resulting in only one nonzero eigenvalue. Since the trace of the full matrix $\gamma_{MF}$ is equal to the particle number which is a basic property of 1-RDMs (depending on the normalization), the trace of $\gamma_{MF}^{\text{env}}$ equates to the particle number without the ones in the fragment. This implies that the one non-zero eigenvalue has to be equal to the number of bosons in the environment thus completing this argument.

\printbibliography

\end{document}